\documentclass[aps,preprint]{revtex4-2}

\usepackage{amsmath}
\usepackage{amssymb}
\usepackage{amsthm}
\usepackage{enumerate}
\usepackage{color}
\usepackage{multirow}
\usepackage{url}
\usepackage{booktabs}
\usepackage{comment}

\newcommand\norm[1]{\left\lVert#1\right\rVert}

\usepackage{tikz}
\usepackage{wrapfig}
\usetikzlibrary{arrows}
\tikzstyle{block}=[draw opacity=0.7,line width=1.4cm]
\usepackage{xspace}

\newcommand{\red}[1]{\textcolor{red}{#1}}

\newcommand{\Z}{\mathbb{Z}}

\newcommand{\R}{\mathbb{R}}

\newcommand{\bs}{\mathbf{s}}

\newcommand{\cC}{\mathcal{C}}
\newcommand{\cF}{\mathcal{F}}
\newcommand{\cG}{\mathcal{G}}
\newcommand{\cK}{\mathcal{K}}

\newcommand{\cP}{\mathcal{P}}
\newcommand{\cR}{\mathcal{R}}

\newcommand{\cV}{\mathcal{V}}

\newcommand{\supp}{\mathtt{supp}}
\newcommand{\Aut}{\mathtt{Aut}}

\newtheorem{definition}{Definition}[section]
\newtheorem{theorem}[definition]{Theorem}

\newtheorem{corollary}[definition]{Corollary}

\usepackage[ruled,vlined]{algorithm2e}

\begin{document}
\title[Symmetry-driven \red{network reconstruction} through pseudobalanced coloring optimization]{Symmetry-driven \red{network reconstruction} through pseudobalanced coloring optimization}

\author{Ian Leifer}
    \affiliation{Levich Institute and Physics Department, City College of New York, New York, NY 10031}
\author{David Phillips}
    \affiliation{Mathematics Department, United States Naval Academy, Annapolis, MD, 21402}
\author{Francesco Sorrentino}
    \affiliation{Department of Mechanical Engineering, University of New Mexico, Albuquerque, NM, 87131}
\author{Hern\'an A. Makse\footnote{Corresponding author: hmakse@ccny.cuny.edu}}
    \affiliation{ Levich Institute and Physics Department, City College of New York, New York, NY 10031}

\begin{abstract}
    Symmetries found through automorphisms or graph fibrations provide important insights in network analysis. Symmetries identify clusters of robust synchronization in the network which improves the understanding of the functionality of complex biological systems. Network symmetries can be determined by finding a {\it balanced coloring} of the graph, which is a node partition in which each cluster of nodes receives the same information (color) from the rest of the graph. \red{In recent work we saw that biological networks such as gene regulatory networks, metabolic networks and neural networks in organisms ranging from bacteria to yeast and humans are rich in fibration symmetries related to the graph balanced coloring.} Networks based on real systems, however, are built on experimental data which are inherently incomplete, due to missing links, collection errors, and natural variations within specimens of the same biological species. \red{Therefore, it is fair to assume that some of the existing symmetries were not detected in our analysis. For that reason,} a method to find pseudosymmetries and repair networks based on those symmetries is important when analyzing real world networks. In this paper we introduce the {\it pseudobalanced coloring} \eqref{eq:mainip} problem, and provide an integer programming  formulation which (a) calculates a pseudobalanced coloring of the graph taking into account the missing data, and (b) optimally repairs the graph with the minimal number of added/removed edges to maximize the symmetry of the graph. We apply our formulation to the {\it C. elegans} connectome to find pseudocoloring and the optimal graph repair. Our solution compares well with a manually curated ground-truth {\it C. elegans} graph as well as solutions generated by other methods of missing link prediction. Furthermore, we provide an extension of the algorithm using Bender's decomposition that allows our formulation to be applied to larger networks.
\end{abstract}

\maketitle

\textbf{Keywords:} Symmetry, Synchronization, Network reconstruction, Missing link prediction, Neural Networks, {\it C. elegans}

\newpage

\tableofcontents

\newpage

\section{Introduction}

{\color{red}
Network models have become a crucial tool in the investigation of biological systems~\cite{Alon19, Buchanan10, Klipp16, StewartNature04}. Examples include neural networks \cite{Varshney11, Bock11}, gene regulatory networks \cite{Guelzim02, Liu15}, metabolic networks \cite{SalmoNet17}, and ecological networks~\cite{PlantPollinator10}. The goal of these studies is to better understand the biological function of the system via a network model. In recent work, Ref. \cite{Morone19} found that {\it automorphisms}~\cite{mckay81} describe the symmetries and function of the neural connectome of the nematode {\it C. elegans} \cite{White86}. An automorphism of a graph is a permutation of nodes that preserves the link structure of the graph (we formally define automorphisms in Section~\ref{sec:intro_and_motivation}). Later \cite{MoronePNAS20, LeiferPLOS20} uncovered fibration symmetries using the algorithms for {\it $K$-balanced coloring} ~\cite{BH,Golubitsky06,mckay81} from \cite{Kamei13, Monteiro21} in biological networks spanning from transcriptional regulatory networks to signaling pathways and the metabolism. Intuitively, a $K$-balanced coloring (the $K$ is omitted when implicit in context) of a graph is a way of partitioning nodes into $K$ clusters such that nodes in the same cluster have the same number of links to every other cluster. This definition can be generalized to the case of weighted links. Section~\ref{sec:definitions} provides a formal definition and a discussion of how graph automorphism, fibration symmetries and balanced coloring are related.

$K$-balanced coloring and resulting fibration symmetries establish a connection between the topology of the graph and its dynamics. We say two nodes in the network are synchronized if they have identical dynamics over time. More precisely, for a network, $G=(V,E)$ made of a set of $V$ nodes, $E$ edges (we use both links and edges to refer to two nodes that are adjacent in a network) and time-varying states, $x_i(t) \in \R^k$ for $i \in V$, we say two nodes $i, j \in V$ are synchronized if $x_i(t) = x_j(t)$ for all time $t$. In this paper we are mainly interested in cluster synchronization formed by two or more synchronized nodes, where several clusters of synchrony can coexist in the network. Complete synchronization, where all the nodes in the network are synchronized in the same state, is not considered in this paper.  As shown in~\cite{Golubitsky06, Deville13, pecora2014cluster, Nijholt16, Morone19, MoronePNAS20, Aguiar21}, the symmetries of the network, given either by automorphisms \cite{pecora2014cluster}, groupoids \cite{Golubitsky06} or graph fibrations \cite{MoronePNAS20}, are associated with patterns of cluster synchrony. Namely, ref.~\cite{Golubitsky06} showed that nodes with the same color in a $K$-balanced coloring, or, in other words, symmetric under the fibration symmetry, form a cluster of synchronized nodes. Furthermore, Leifer {\it et al.}~\cite{LeiferBMC20} used available experimental data on gene co-expression in bacteria to confirm the existence of cluster synchronization predicted solely by the fibration symmetries in these networks. Existence of these synchronous solutions can considerably augment our understanding of the functionality of the modeled system \cite{MoronePNAS20, LeiferPLOS20}.

Symmetry and synchronization are important concepts in the field of physics. Symmetry lays in the foundation of the standard model and has broad applications in other parts of physics from Lagrangian mechanics to crystallography and nuclear spectroscopy. One of the fundamental problems in chaos theory is the search for synchronous solutions in dynamical systems \cite{Strogatz18}.  In the context of biological networks, in particular in neural networks, complete synchronization is often studied by considering the dynamics of networks of coupled oscillators using the Kuromoto model \cite{Arenas08, Rodrigues16}. Methods of statistical mechanics have broad applications in the studies of networks of coupled oscillators \cite{Rodrigues16, Pikovsky01, Gupta14}. For example, the transition of a network of coupled oscillators to the synchronous state, called the synchronization transition, can be thought of as a phase transition (bifurcation) in the ensemble of nodes of the network of oscillators \cite{Gupta14}, which then permits using the tools of statistical mechanics developed for study of phase transitions.
}

Despite these advances, modeling biological systems as networks with symmetries pose additional challenges. Firstly, experimental data on networks collected by different techniques are never complete due to experimental errors and the impossibility to measure every possible link. The missing links include protein-protein interactions, binding between transcription factors and DNA, neural synaptic connections in the brain or even metabolic reactions.  Secondly, natural variability across individuals of the same species results in different networks making interpretation of the data across specimens more difficult \cite{Varshney11}. 
Thirdly, organisms adjust to the changing habitat using neural plasticity, epigenetics and other adaptations. For example, even two organisms that were identical at a given moment of time may become different in the future. Organisms survive and benefit from these small variations, so evolution through natural selection occurs. Therefore, networks that model these systems exhibit not only differences across individuals, but also have missing experimental links that can mask their underlying regularities. In particular, the missing experimental interactions and natural variations make the existence of perfect symmetries difficult to find in biological networks. Thus, symmetries in biological networks have been very hard to find since the time of Monod's first formulation of the problem \cite{Monod70}. The difficulty persisted despite the ubiquitous existence of synchronization across all biological networks, which is a manifestation of symmetries in the underlying networks supporting the biological activity \cite{LeiferBMC20}.

A typical case study is the connectome of the nematode {\it C. elegans}, which was fully mapped in a landmark paper in 1986 \cite{White86}. Despite of being one of the simplest connectomes, containing only 302 neurons in the hermaphrodite, there is plenty of variability from animal to animal (about 25\% of links are different between animals \cite{Varshney11}) and modern measurements and data curation \cite{Varshney11} keep finding new synaptic connections that were missed before \cite{White86}.
Despite this inherent disorder, the function and neural activity of {\it C. elegans} exhibits strong regularities, as observed, for instance, in the oscillatory locomotion patterns and the neural synchronization observed in neuronal activity \cite{Kato15}.  

What is the anatomical substrate for this synchrony? Theory predicts that robust synchronization arises from the symmetries in the underlying network. 
However, perfect symmetries are impossible to be realized in biology. Therefore, Morone {\it et al.} \cite{Morone19} introduced the concept of \textit{pseudosymmetries} which are almost-symmetries of the incomplete individual network that, upon reconstruction of a few missing links, they reveal an underlying ideal perfect symmetry that is the unique "blueprint" of the symmetry of the species. 
Each individual network can be thought of as a small variation of this "blueprint" network containing the ideal symmetries. The observed experimental networks are all different and always pseudosymmetric and can be slightly modified to exhibit the perfect "blueprint" of the species.

In this paper we introduce the {\it pseudobalanced $K$-coloring}  problem, and an integer programming formulation \eqref{eq:mainip} to find the minimum number of missing links in the network to achieve symmetry.  We are interested in biological networks since they are the substrate for cluster synchronization of their units. These include brain networks (connectomes), gene regulatory networks, metabolic networks and others. While the symmetries are not perfect, yet, they are close enough to ideal symmetries such that they can sustain the observed experimental cluster synchronization and functional regularities. We present an optimal integer linear programming formulation of the problem that attempts to identify the pseudosymmetries of the network based on pseudobalanced coloring, and at the same time, reconstruct the network with a minimum number of modified links to transform the pseudosymmetric network into ideal symmetry. Beyond biological networks, our symmetry-driven reconstruction can be applied to any network that supports cluster synchronization of its units.


Reconstructing a network to reveal ideal symmetries is an instance of the problem of link prediction in complex networks~\cite{Lu11}. Link prediction is the problem of determining the probability of the existence of a link between a pair of nodes in the network that is missing from the data \cite{Lu11, Getoor05, Liben07, Chen05, Newman08}. Previous work proposed several approaches to find missing links based on different statistical and Markov chain models, see review in \cite{Lu11}. However functional and dynamical features of the incomplete networks, such as  synchronizability, are rarely taken into consideration. The approach presented here employs network symmetry as an organizing principle of the network structure to guide the link prediction and reconstruct the network from its incomplete state.

\begin{figure}[ht!]
	\centering
	\includegraphics[width=0.8\linewidth]{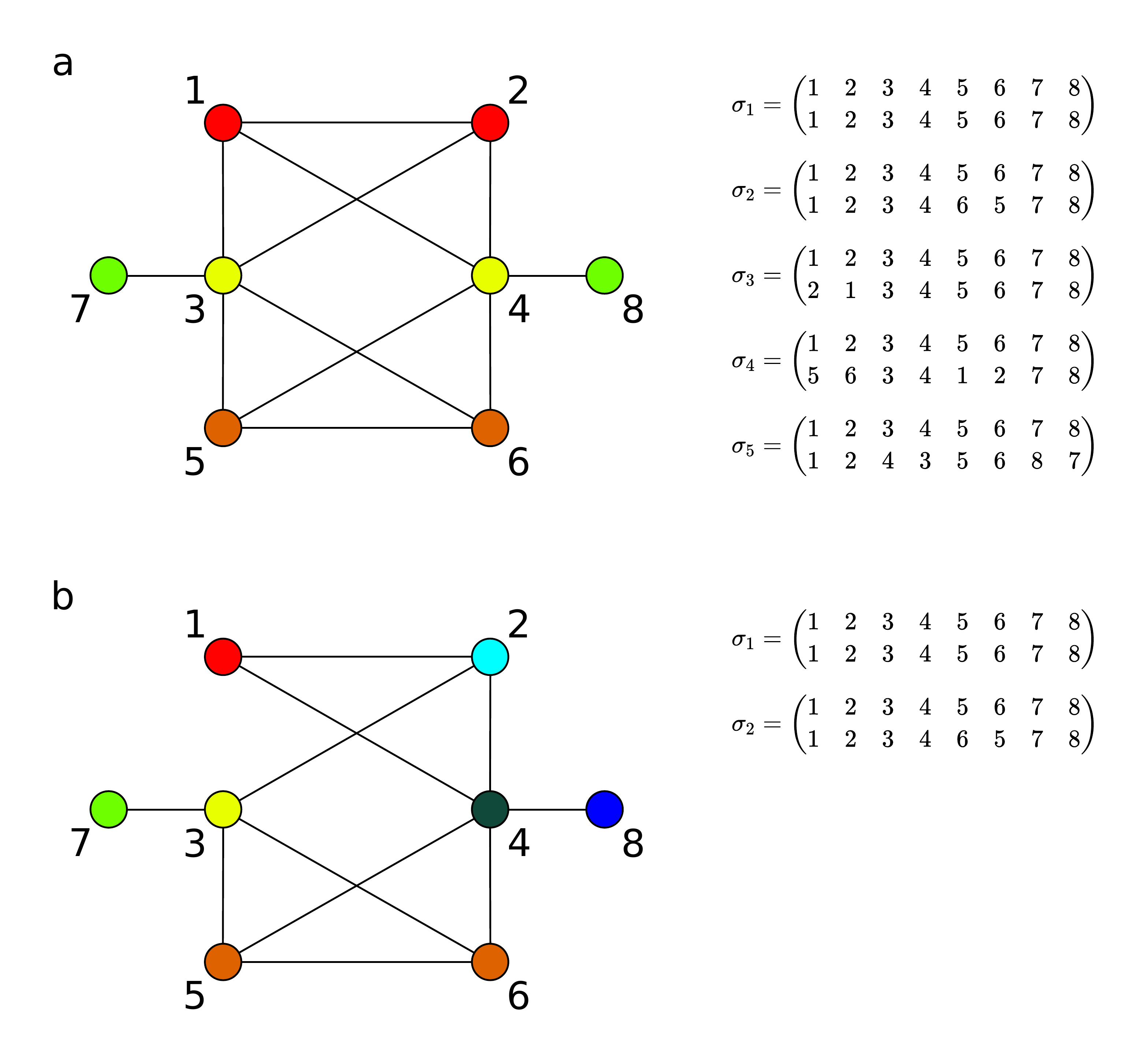}
	\caption{Differences between automorphism groups and balance coloring of two similar graphs (a) and (b) differing by only one link $(1, 3)$. The automorphism group of graph (a) consists of 5 elements: $\sigma_1, \dots, \sigma_5$ whereas the automorphism group of graph (b) consists of 2 elements: $\sigma_1$ and $\sigma_5$. Graph coloring is balanced in both networks, i.e. nodes of the same color have the same number of links to every other color, but graph (a) is much 'more symmetric' than (b) and requires a fewer non-trivial colors. On the contrary, missing a single link makes the graph in (b) less symmetric as seen from the abundance of trivial colors (colors assigned to just one node) needed to color the graph. The purpose of this work is to develop a formulation that would identify this missing link that makes the graph more symmetric, in a precise way.}
	\label{Fig1}
\end{figure}

Traditionally, network symmetry is defined by using graph automorphisms or symmetry permutations. An automorphism is a permutation of nodes that preserves the adjacency structure of the graph ~\cite{Harary, Dixon96}. That is, before and after the permutation the nodes are connected to the same nodes. The set of all automorphisms of a graph form the symmetry group of the graph. Automorphisms are, however, sensitive to small perturbations in the graph, i.e., the removal of one link can drastically change the automorphism groups~\cite{Morone19, MoronePNAS20, YanchenApprox20}. For an explicit example, consider the two graphs in Fig.~\ref{Fig1} and their corresponding automorphisms presented on the right side. Graphs in figures \ref{Fig1}a and \ref{Fig1}b differ only by link $(1, 3)$. Missing  this link has a global effect on the symmetry of the network decreasing the size of the automorphism group from five elements ($\sigma_1, \dots, \sigma_5$) to only two ($\sigma_1$ and $\sigma_2$). Therefore, the symmetry of the graph can be substantially reduced by missing just a few links in the topology of a graph. Inversely, the graph in Fig.~\ref{Fig1}b can be restored to be a perfectly symmetric graph in Fig.~\ref{Fig1}a by adding just one link.

The automorphisms of the graphs are important since they lead to robust cluster synchronization of the nodes activity. 
The symmetry group of the network leads to synchronization of the nodes associated by the orbits of the graph. An orbit of a node is the set of nodes that are obtained by the application of all the automorphisms of the graph. The orbits are non-overlapping clusters and each orbit can sustain an independent cluster synchronization, although its stability is not guaranteed. 

The pseudosymmetry formulation proposed in \cite{Morone19} attempts to reconstruct the graph 
based on the automorphism group of the network. However, automorphisms cannot predict all the possible symmetries and cluster synchronization present in the network. Instead, balanced coloring~\cite{Golubitsky06} and associated symmetry fibrations \cite{MoronePNAS20, LeiferPLOS20} provide a more general type of symmetry. Since all orbits of the graph are also balanced colored, but not the opposite \cite{Kudose09}, a balanced coloring-based formulation captures more cluster synchronizations than those predicted by automorphisms.
In this paper,  we 
derive the symmetry reconstruction of the graph by finding the ideal balanced coloring of the graph rather than the ideal automorphisms as done in \cite{Morone19}. Thus, we will reconstruct the network based on {\it pseudobalanced coloring}. Reconstructing the network has two objectives: (1) determine the "best" number of clusters, i.e., determine $K$, and (2) perform the minimum number of {\it perturbations} to perform on the graph. In general, we use perturbation to refer to any change to a link weight with the convention that reducing a link weight to zero removes the link altogether. For this paper, we focus on the case of unweighted links and only adding links to the graph. In particular, we are interested in the following optimization problem: find the minimum number of new links to add to a graph so that it possesses a balanced $K$-coloring.  After solving this problem for the allowable values of $K$, we use graph functions to determine the "best" choice of $K$.


A balanced coloring with the minimal number of colors gives rise to a symmetry fibration, which is a transformation that collapses all the colors of the graph into a representative node of the base. Thus, the present formulation based on  pseudobalanced coloring is related to finding a quasi-fibrations of the graphs \cite{Boldi21}. Ref. \cite{Boldi21} deals with this analogous graph reconstruction formulation based on quasi-fibrations.

For smaller networks, adding links manually to find symmetry is possible \cite{Morone19}, but such a method is computationally impractical for larger networks. Moreover, such ad hoc methods do not have an optimality guarantees. Thus, an important endeavor is to develop models and algorithms that can find symmetry in the presence of small perturbations in the underlying input data. In this paper, we describe an integer programming formulation that finds the minimum number of edges to add to a given network so it has a balanced coloring~\cite{BH} of a given size (balanced colorings are also referred to as equitable partitions~\cite{mckay81}). \red{Our complexity results in Section~\ref{sec:complexity} show that the directed variant of our problem is weakly NP-Hard and we conjecture the same is true of the undirected version so an efficient algorithm is unlikely to exist for our problems of interest.} Therefore, integer programming is a natural method to use to solve our problem. We are able to solve the integer program for modestly sized instances (20-30 nodes and 100-120 edges) of interest using Gurobi 9.11~\cite{gurobi}. \red{Most instances required seconds to solve although there were cases where hours were needed. In preliminary tests, we found that solving larger instances with approximately 280 nodes and 3000 edges required minutes to hours to solve. We emphasize that we do not draw conclusions from these results as we did not conduct a full computational study. For future work, we also describe an algorithm based on a Bender's decomposition of the integer program in Appendix~\ref{sec:benders} to help improve the computational efficiency of solving the integer program for larger scale instances.}

The paper is organized as follows. In Section \ref{sec:intro_and_motivation}, we review the previous pseudosymmetry formulation \cite{Morone19} and provide intuition and motivation for our method. In Section \ref{sec:definitions}, we give formal definitions for the problems of interest. In Section \ref{sec:complexity}, we prove that a simplified variant of our problem is weakly NP-Hard. In Section \ref{sec:formulation}, we describe the integer linear programming formulation of our problem. In Section \ref{sec:results}, we present results from our integer linear programming approach applied to experimental data on the {\it C. elegans} connectome. In Section \ref{sec:normal_subgroup_decomposition}, we show how a repaired graph can be partitioned into co-functioning clusters. In Section \ref{sec:comparison_of_methods}, we compare results obtained by using different objectives with some of the traditional link prediction methods. In Section \ref{sec:conclusion}, we discuss implications from our results and outline future work. Appendix \ref{sec:benders} provides a Bender's decomposition and algorithm for our integer linear program.

\section{Previous work on pseudosymmetry}
\label{sec:intro_and_motivation}

Consider for the moment an undirected graph $G=(V,E)$ and let $\pi$ denote a permutation of the node labels, i.e. $\pi: V \rightarrow V$. Then $\pi$ is an automorphism if, and only if, for all $u, v \in V$, if $uv \in E$ then $\pi(u)\pi(v) \in E$. The automorphism group, denoted $\Aut(G)$, is the set of all automorphisms of $G$ along with function composition as the multiplication. A useful characterization of automorphisms involves permutation matrices.
Let $A$ be a node-node adjacency matrix of $G$ and $P$ denote the permutation matrix associated with a permutation $\pi$. Then $\pi \in \Aut(G)$ if and only if
\begin{equation}
\label{eq:perm}
  [P, A] = PA - AP = \mathbf{0}
\end{equation}
where $\mathbf{0}$ denotes the matrix of all zeros. Let $\cP$ denote the set of $|V| \times |V|$ matrices. If the permutation matrix $P \in \cP$ satisfies Eq.~\eqref{eq:perm}, we say that $P$ is a {\it perfect symmetry}. The {\it trivial symmetry} is the identity matrix and is shared by all graphs.

In order to study graphs that slightly vary from those with perfect symmetry,
Morone and Makse \cite{Morone19} suggest looking for pseudosymmetries defined by a small parameter $\varepsilon\in \R$:
\begin{equation}
    \label{Eq:psedosymm_natcomm}
    \cP_\varepsilon = \{P_\varepsilon \in \cP: \norm{[P_\varepsilon, A]} \leq \varepsilon\}
\end{equation}
where $||X||$ is the Frobenius norm of the matrix $X = (x_{ij})$, i.e., $||X|| = \sqrt{\sum_{i = 1, j = 1}^{n, m}x_{ij}^2}$. For $\varepsilon = 0$ this definition is the same as perfect symmetry. Note, the set of permutations associated with $\cP_\varepsilon$ doesn't necessarily form a group. The naive way to find elements of $\cP_\varepsilon$ is to consider all possible permutations and determine whether Eq.~\eqref{Eq:psedosymm_natcomm} is satisfied. Because $|\cP| = (|V|)!$ this naive approach is computationally intractable and finding a more efficient method is desirable. Moreover, for a given  permutation matrix that satisfies Eq.~\eqref{Eq:psedosymm_natcomm}, we would like a method that finds a related graph that is symmetric. Thus, we want an alternative approach that also finds a way to ``repair" $G$, i.e., minimally perturbs $G$ and results in a symmetric permutation matrix.

Because the Frobenius norm is unitarily invariant i.e. $\norm{X} = \norm{UXV}$ for all matrices $X$ and unitary matrices $U$ and $V$, 
\begin{equation}
    \norm{[P_\varepsilon, A]} = \norm{P^\varepsilon  A - A  P^\varepsilon} = \norm{P^\varepsilon  A  (P^\varepsilon)^{-1} - A}. 
    \label{Eq:psedosymm_natcomm2}
\end{equation}

We focus on the case of adding edges to an undirected graph $G$ and show that, in this case, $\varepsilon$ is bound by four times the number of added edges. Suppose that we have a graph, $G$, with added edges so that $P^\varepsilon$ is a symmetry and let the adjacency matrix of this graph be denoted by $A_\varepsilon$. Then $P^\varepsilon  A_\varepsilon (P^\varepsilon)^{-1} - A_\varepsilon = \mathbf{0}$, equation \eqref{Eq:psedosymm_natcomm2}, and the triangle inequality imply that:
\begin{equation}
    \begin{aligned}
        \norm{[P_\varepsilon,A]} & = \norm{A - P^\varepsilon_i  A  (P^\varepsilon_i)^{-1}} &  \\
         & = \norm{P^\varepsilon_i  A_\varepsilon  (P^\varepsilon_i)^{-1} - P^\varepsilon_i  A  (P^\varepsilon_i)^{-1} - A_\varepsilon + A} &  \\
         & = \norm{P^\varepsilon_i  (A_\varepsilon - A)  (P^\varepsilon_i)^{-1} - (A_\varepsilon - A)} & \\
         & \leq \norm{P^\varepsilon_i  (A_\varepsilon - A)  (P^\varepsilon_i)^{-1}} + \norm{(A_\varepsilon - A)}  & \\
         & = 2 \norm{A_\varepsilon - A}. &
    \end{aligned}
\end{equation}

The difference matrix $(A_\varepsilon - A)$ is the adjacency matrix corresponding to the graph that just has the repaired edges. Therefore, $\norm{[P_\varepsilon,A]}$ is less than the number of repaired edges multiplied by 4 (2 because the graph is undirected and $A$ is symmetric and another 2 because of the coefficient in front of the norm). Hence, a reasonable approach to repair $G$ is to add the minimum number of edges required for the graph to have a symmetry. 

To summarize this approach, finding pseudosymmetries in a graph using brute force is computationally intractable and not guaranteed to find an optimally repaired graph with minimal modifications. Instead, in this paper, we formulate the problem in terms of the balanced coloring problem. It is known that balanced coloring is a pre-processing step in finding the automorphisms of the network as it is used in the popular McKay's algorithm Nauty \cite{mckay81}.  This is  because all orbits are balanced colored (but not the opposite). Thus, finding first a minimal balanced coloring of the graph (which can be found in quasilinear time) leads to the orbits of the graphs, which can then be used to find the generators of the automorphism group.  In Section~\ref{sec:definitions}, we make this idea more precise.

Based on these ideas, rather than searching for pseudo automorphisms like in \cite{Morone19}, here we search for pseudobalanced coloring \cite{MoronePNAS20,LeiferPLOS20}, which are more general symmetries than automorphisms. We use an integer linear program that adds the minimum number of edges to a graph so that we make the graph "more symmetric" in terms of balanced coloring. Figure \ref{Fig1}a shows the perfect balanced coloring (see definition in next section) found in the symmetric network and the comparison with the same perfect balanced coloring applied to the incomplete network in Figure \ref{Fig1}b. We see how a single missing link can destroy not only the automorphism group but also the perfect balanced coloring. The goal of the formulation is to first find the pseudobalanced coloring of Fig. \ref{Fig1}a in the graph in Fig. \ref{Fig1}b, and at the same time repair the missing link that transform the pseudobalanced coloring into a perfect balanced coloring for Fig. \ref{Fig1}b.


\section{Definitions and problems}
\label{sec:definitions}
In this section, we describe both a general version of the pseudobalanced $K$-coloring problem and the specific case that we are interested in. We believe the general version contains many interesting variants that pose important challenges to be solved as the graph can be undirected or directed and weighted or unweighted. In addition, the perturbations on the graph in the most general version are permitted to have very general restrictions or lack thereof. The specific version of our problem is on an undirected, unweighted graph and the only perturbations allowed are the addition of edges.

Let $G=(V,E)$ be a given graph. We let $A \in \R^{|V| \times |V|}$ denote the weighted node-node adjacency matrix associated with $G$. We also define the {\it directed} complement of $E$ to be $E^C = \{(i,j) \in V \times V: ij\not\in E, i\not=j\}$ and the undirected complement of $E$ to be $E' = \{ij: i,j \in V, ij \not \in E\}$. We require both $E^C$ and $E'$ for our formulation when the graph is undirected. When the graph is directed, $E^C = E'$. To help emphasize the graph type, we use the convention that consecutive node indices, e.g., $ij$, represent an undirected edge. An ordered pair of node indices, e.g., $(i,j)$, represent a directed edge. We also recall that a {\it partition} of a given set $S$, is a collection of pairwise disjoint subsets of $S$ whose union is all of $S$, i.e., if $\cC$ is a partition of $S$ then
\begin{equation}
    \label{eq:partition}
    \bigcup_{C \in \cC} C = S \,\, \mbox{ and,}\,\,
    \mbox{ for all $C,D \in \cC$, } \,\, C \cap D = \emptyset.
\end{equation}

Balanced coloring has been defined by several authors, e.g., see \cite{Mckay76,mckay81,mckay14,BH} for a definition corresponding to the one we use where it is sometimes referred to as an equitable  partition.
A stricter definition of balanced coloring is given and used in \cite{Kamei13,MoronePNAS20} although their definition is the same as ours for unweighted, i.e., binary graphs. There are different definitions including ones that are unrelated and more similar to traditional graph coloring, e.g., \cite{Feige10}.

In our definition, we fix $K$, the number of colors, i.e., clusters, as the objectives of our eventual problem are to both determine the minimum number of links to add to a graph so that a balanced coloring exists, but also to determine the "best" number of colors, i.e., the best $K$. 
\begin{definition} {\bf Balanced $K$-coloring.}
  A {\it balanced $K$-coloring} of $G$ is a partition, $\cC$, of the node set $V$ which satisfies the following: 
\begin{itemize}
    \item The cardinality of $\cC$ is $K$.
    \item In the case that $G$ is an undirected graph, the condition is as follows. For all $C \in \cC$, all pairs of distinct nodes $p,q \in C$, and all $D \in \cC$,
\begin{equation}
  \label{eq:balanced}
  \sum_{j \in D: pj \in E} A_{pj} = \sum_{j \in D: qj \in E} A_{qj}.
\end{equation}
    \item In the case that $G$ is a directed graph, Eq.~\eqref{eq:balanced} corresponds to two separate conditions resulting in three kinds of balanced coloring. The two conditions are as follows. For all $C \in \cC$, all pairs of distinct nodes $p,q \in C$, and all $D \in \cC$,
\begin{equation}
  \label{eq:balanced-out}
  \sum_{j \in D: (p,j) \in E} A_{pj} = \sum_{j \in D: (q,j) \in E} A_{qj}
\end{equation}
and
\begin{equation}
  \label{eq:balanced-in}
  \sum_{j \in D: (j,p) \in E} A_{jp} = \sum_{j \in D: (j,q) \in E} A_{jq}.
\end{equation}
\end{itemize}
A {\it directed out-balanced coloring} is if only Eq.~\eqref{eq:balanced-out} is enforced and a {\it directed in-balanced coloring} is if only Eq.~\eqref{eq:balanced-in} is enforced. A {fully directed balanced $K$-coloring} has both conditions enforced. 
\end{definition}
\red{We also define the {\it minimal balanced coloring} which, intuitively,} corresponds to the minimum number of colors $K$ to color the graph so that no nodes with different colors are balanced. Because the number of colors, $K$, to color a graph is unique~\cite{Stewart07}, we omit $K$ \red{in this definition.}
\begin{definition} {\bf Minimal balanced coloring.}
A {\it minimal balanced coloring} is a balanced $K$-coloring, $\cC$, where
the following is also true. For every $p \in V$, let $C_p \in \cC$ be
the unique set that contains $p$. For every distinct $p, q \in V$
such that $q \not\in C_p$ there exists $D
\in \cC$ such that Eq.~\eqref{eq:balanced} is violated for the undirected case or Eq.~\eqref{eq:balanced-in} or Eq.~\eqref{eq:balanced-out} is violated in the directed case.

 For example, in the undirected case, the following must be true for some set $D \in \cC$.
\begin{equation}
  \label{eq:minimalbalanced}
  \sum_{j \in D: pj \in E} A_{pj} \not= \sum_{j \in D:qj \in E} A_{qj}.
\end{equation}
\end{definition}

In other words, in an in-balanced coloring partition, two nodes with the same color 'receive' the same colors from connected nodes. A minimal balanced coloring is a balanced coloring with the minimal number of colors. We call a color \textit{trivial} if there is only one node that belongs to this color. A \textit{non-trivial color} is a color that is not trivial. A coloring in which each color is trivial (corresponding to the discrete partition) is called discrete. We refer to a node as {\it trivial} if it possesses a trivial color and call a node {\it symmetric} if its color is non-trivial. Abusing this definition, we also say a node is symmetric to another if they are both the same color. 

\underline{Relation to graph fibrations.} In the framework of graph fibrations \cite{MoronePNAS20}, a cluster of nodes with the same \red{balanced} color is \red{equivalent to} a {\it fiber} \red{(see Methods in ref.~\cite{LeiferBMC20})}.  Nodes in a fiber have isomorphic input trees. When an admissible set of dynamical equations is attached to the graph, nodes of the same balanced color (or fiber) are predicted to be synchronous (cluster synchronization). A symmetry fibration is a transformation that collapses nodes in a minimal balanced color cluster into a single representative node in the base of the graph. We note that, while nodes belonging to the same orbit are also synchronous, nodes of the same balanced color do not necessarily belong to the same orbit of the automorphism group. However, all nodes in each orbit do have the same balanced color \cite{mckay21web, Kudose09}. Thus, balanced coloring, fibers and symmetry fibrations represent more general symmetries than automorphisms, and reveal cluster synchronization in the graph that is not captured by orbital partitions,  see \cite{MoronePNAS20} for more details. 

\underline{Relation to graph automorphisms.} The minimal balanced coloring and automorphisms are fundamentally related. In this case, the number of colors found is a lower bound on the number of automorphism orbits. Moreover, the nodes in each nontrivial color found corresponds to potential automorphisms of the graph. 

To explain the connection between coloring and automorphisms further, we describe how graph automorphisms and isomorphisms are found. (Note that the two problems are equivalent~\cite{GareyJ78}) The graph isomorphism and automorphism algorithms we are aware of \cite{Mckay76,Darga04,Junttila07,Piperno08,Lopez09} all use a search tree to find isomorphisms/automorphisms.  For a given graph $G$, each node of this search tree represents a balanced coloring of $G$. The root of the tree corresponds to the minimal balanced coloring. Each child in the search tree is found using the \textit{individualization-refinement} process. In this process a child's coloring is obtained by \textit{refining} its parent's coloring by choosing a node of a non-trivial color (in a parent), assigning it with a unique color and obtaining a new balanced coloring of the graph while keeping this node \textit{individualized} (having a unique (trivial) color). This individualization-refinement process is repeated until the coloring is discrete, hence all leaves of the search tree correspond to the discrete coloring. In the last step, strings corresponding to all leaves are constructed by combining the rows of the adjacency matrix in the order defined by the coloring of each leaf. Due to the fact that the search tree is isomorphism-invariant, whenever strings corresponding to two leaves are the same, colored graphs in these two leaves are isomorphic and hence a color-preserving mapping between these two graphs is an automorphism of $G$. Readers interested in more details can refer to \cite{mckay14, Piperno08, Grohe17, mckay21web}.

The size of graph's search tree has a deep connection with the number of non-trivial colors in the minimal balanced coloring of the graph. Each non-trivial color requires refinement resulting in more children nodes and thereby increasing the search tree size. Therefore, repairing the graph to a version with similar topology but  fewer nodes trivially colored so that more nodes have non-trivial colors than in the original graph. Repairing graphs in this way creates a "more symmetric" version of the graph, that is, a version of the graph that has (or at least may have) more elements in the automorphism group. Any graph could be be repaired to a complete graph, i.e., a graph with a link between every pair of nodes. The automorphism group of the complete graph is a symmetric group, but naturally is unlikely to be similar topologically to the original graph. In particular, we wish to use the minimum number of repairs necessary to find a more symmetric version of the graph so that the essential network topology of the original graph is maintained as much as possible. Thus, the repairing process must balance the trade-off between making the graph more symmetric (by increasing the number of non-trivial colors) versus making minimal changes to the graph topology.

{\color{red}
\underline{Relation to stochastic block models.} It is worth mentioning a parallel to stochastic block models (SBMs) introduced in the context of social networks by Holland \textit{et al.} \cite{Holland83}. SBM is a graph generative model in which the graph is partitioned into groups and edges between nodes are placed with the probability defined by the clusters these nodes belong to. In other words, SBMs are the stochastic version of a graph partition (different from the equitable partition), where the analogous constraints of Eq. 6, 7 and 8 hold only in \emph{expectation}. We refer readers to ref.~\cite{Schaub20} (section IIA and figure 4) for a more rigorous discussion of the connection between SBMS and equitable partitions (i.e., balanced coloring, fibers).
}

We now define the general version of this problem. In this version, existing edge weights are allowed to be perturbed, i.e., changed, and non-existent edges are allowed to be introduced with weights restricted in some manner (or not). We still wish for a minimum amount of perturbations so that the balanced conditions are enforced. We first formally define the set of {\it generalized pseudobalanced colorings} of a graph and then the corresponding optimization problem which occurs over this set. \red{We also add the option that the colorings considered must adhere to some prior knowledge about the given graph. In particular, we allow that some nodes' colors are already known. Such an option corresponds to prior expert knowledge about the graph structure and our methods permit this restriction.}

\begin{definition}
\label{eq:general-perturbation}
{\bf Generalized pseudobalanced $K$-coloring}. Let $G=(V,E)$ denote a graph with edge weight matrix $A \in \R^{|V| \times |V|}$ and $\cR \subseteq \R^{|E|+|E'|}$ given. \red{Let $\cF$ denote a collection of disjoint subsets of $V$.} A generalized pseudobalanced coloring of $G$ is an ordered triple $(\cC,\Phi, \Omega)$ with the following true.
\begin{itemize}
    \item $\cC$, the coloring, is a partition of the node set $V$ with $|\cC| = K$, i.e, $\cC$ satisfies Eq.~\eqref{eq:partition}.
    \item \red{If $\cF \not=\emptyset$ then for all pair of distinct sets $C, D \in \cF$, there must exist a pair of distinct sets $S,T \in \cC$ such that $C \subseteq S$ and $D \subseteq T$. If $\cC$ satisfies this condition, then we say that $\cC$ {\it respects} $\cF$.}
    \item $\Phi = (\phi_{ij}) \in \R^{|E|}$ is a vector of perturbations on the edge weights.
    \item $\Omega = (\omega_{ij}) \in \R^{|E'|}$ is a vector of new weights that do \underline{not} exist in $G$. 
    \item The perturbation vectors $\Phi$ and $\Omega$ are related and restricted through the set $\cR$, i.e., 
    \begin{equation}
        \label{eq:perturb-restriction}
        (\Phi, \Omega) \in \cR.
    \end{equation}
    \item Appropriate balanced equations are satisfied, i.e., Eq.~\eqref{eq:balanced}, Eq.~\eqref{eq:balanced-in}, and/or Eq.~\eqref{eq:balanced-out} are satisfied by $(\cC,\Phi,\Omega)$ depending on whether $G$ is undirected or directed. If undirected, then for all $C \in \cC$, all pairs of distinct nodes $p, q \in C$ and all $D \in \cC$,
\begin{equation}
  \label{eq:pseudobalanced}
  \sum_{j \in D:pj \in E} (A_{pj}+\varphi_{pj}) + \sum_{j \in D:pj \in E'}
  \omega_{pj} = \sum_{j \in D:qj \in E} (A_{qj}+\varphi_{qj}) + \sum_{j \in D: qj
    \in E'} \omega_{qj}.
\end{equation}
If $G$ is directed, then one or both of the following two conditions replaces Eq. \eqref{eq:pseudobalanced}. Recall that $E$ and $E'$ are ordered pairs of nodes in what follows. For all $C \in \cC$, all pairs of distinct nodes $p, q \in C$ and all $D \in \cC$,
\begin{equation}
  \label{eq:pseudobalanced-in}
  \sum_{j \in D:(p,j) \in E} (A_{pj}+\varphi_{pj}) + \sum_{j \in D:(p,j) \in E'}
  \omega_{pj} = \sum_{j \in D:(q,j) \in E} (A_{qj}+\varphi_{qj}) + \sum_{j \in D: (q,j) \in E'} \omega_{qj}
\end{equation}
\begin{equation}
  \label{eq:pseudobalanced-out}
  \sum_{j \in D:(j,p) \in E} (A_{jp}+\varphi_{jp}) + \sum_{j \in D:(j,p) \in E'}
  \omega_{jp} = \sum_{j \in D:(j,q) \in E} (A_{jq}+\varphi_{jq}) + \sum_{j \in D: (j,q) \in E'} \omega_{jq}
\end{equation}
\end{itemize}  
For a given positive integer $K \in \{1,\ldots,|V|\}$, we let $\cG_{G,K,\cF,\cR}$ denote the set of all generalized pseudobalanced $K$-colorings on $G$ with fixed nodes $\cF$, i.e., $|\cC| = K$, \red{$\cC$ respects the subsets of $\cF$,} and  edge and non-edge perturbations restricted to $\cR$. In particular,
\[
\cG_{G,K,\cF,\cR} = \left\{(\cC,\Phi,\Omega): \red{\cC \mbox{ respects } \cF}, |\cC|=K, \eqref{eq:partition}, \eqref{eq:perturb-restriction}, \mbox{ and } (*) \mbox{ are satisfied.} \right\}
\]
Here, $(*)$ represents the appropriate choice(s) of \eqref{eq:pseudobalanced}, \eqref{eq:pseudobalanced-in}, and/or \eqref{eq:pseudobalanced-out}.
\end{definition}

\begin{definition} {\bf The Optimal Repair Problem (ORP).}
The optimization problem of interest is then find the minimum amount of perturbations to the edges (addition/removal/weight changes) so that a balanced $K$-coloring exists on the perturbed graph:
\begin{equation}
\label{eq:generalized-opt-pc}
\min\left\{ \sum_{ij \in E} |\omega_{ij}| + \sum_{ij \in E'} |\phi_{ij}| :
(\cC,\Phi,\Omega) \in \cG_{K,G,\cF,\cR} \right\}.
\end{equation}
\end{definition}
\noindent
There are many interesting cases of ORP, Eq.~\eqref{eq:generalized-opt-pc}, involving the particular restrictions to $\cR$, e.g., $\cR$ are box constrained by an appropriate small constant or $\cR$ is an elliptical set. In this paper, we focus on the following restricted case:  We consider undirected and unweighted graphs where links are permitted to be added but not removed from the graph.  We note, however, that our methods can be extended to directed graphs and also weighted edges. 

We call a {\it restricted pseudobalanced $K$-coloring} of a given graph as the case of the generalized pseudobalanced $K$-coloring where the graph is unweighted, link perturbations and removals are not allowed, and only new unweighted links are allowed to be added. \red{We also allow for some of the nodes to have fixed colors in advance, e.g., by running a balanced coloring algorithm in advance of repairs or leveraging expert knowledge.} Formally, we define this as follows and we call it {\it pseudobalanced $K$-coloring} for short:

\begin{definition}{\bf Pseudobalanced $K$-coloring (PBC).}
  \label{def:pseudo}
    Let $G=(V,E)$ be a given undirected graph with node-node adjacency matrix $A \in \{0,1\}^{|V| \times |V|}$, and $K \in \{1,\ldots,|V|\}$ and $\cF$ to be a given collection of disjoint subsets of $V$. Denote the set of pseudobalanced $K$-colorings of $G$ by $\cP_{G,K}$ where
    \[
    \cP_{G,K,\cF} = G_{G,K,\cF,\{0\}^{|E|} \times \{0,1\}^{|E'|}},
    \]
    i.e., $\phi_{ij} = 0$ for all $ij \in E$ and $\omega_{ij} \in \{0,1\}$ for all $ij \in E'$.
    The optimization problem Eq.~\eqref{eq:generalized-opt-pc} reduces to
    \begin{equation}
    \label{eq:sumPseudo}
    \min\left\{\sum_{i,j \in V, ij \not\in E} \omega_{ij} : (\cC,\Omega) \in \cP_{K,G,\cF}, 
    \forall \omega_{ij} \in \Omega\right\}.
\end{equation}
\end{definition}
As we show in the subsequent section, \red{solving the in-balanced ORP on directed graphs is unlikely to be efficiently solvable. We conjecture the same is true of the (PBC).} 

\section{Problem complexity}
\label{sec:complexity}
The first step of many algorithms for graph isomorphism is to find the minimal balanced coloring, which is polynomial time solvable~\cite{Mckay76,Darga04,Junttila07,Piperno08,Lopez09}.  In our paper, we use the algorithm of Kamei and Cock \cite{Kamei13}. The complexity of the variants we have described of the balanced coloring problem are, to the best of our knowledge, unknown. In this section, we prove that \red{finding an in-balanced $K$-coloring on a weighted graph is weakly NP-Hard when $\cF \not= \emptyset$. We note that the four conditions that the proof relies on are (1) the restriction that exactly $K$ colors must be used, (2) the allowance of weighted edges, (3) that the graph is directed, and (4) the inclusion of fixed nodes, i.e., $\cF \not= \emptyset$. As a corollary, the in-balanced ORP on directed weighted graphs is also weakly NP-Hard.}

\red{Our proof reduces the weighted, directed in-balanced $K$-coloring problem to the partition problem. In this problem, a list of positive integers is given and the decision problem is to determine if there is a way to partition the list into 2 sets elements so that the sums of each set is equal. Formally, this problem can be stated as follows.}
\begin{definition}
\label{def:partition} 
\red{Let $n \in \Z_+,$ and a set of positive integers, denoted $X = \{x_1, \ldots, x_{n}\}$, be given whose sum is even.
The decision problem is to determine if there is a partition of $X$ into $2$ sets, denoted by $S$ and $T$ so that
\begin{equation}
    \label{eq:partition_balance}
\sum_{x \in S} x = \sum_{x \in T} x. 
\end{equation}
}
\end{definition}
\noindent 
\red{The partition problem is weakly NP-Hard~\cite{GareyJ78} and we can show that being able to solve the weighted directed in-balanced $K$-coloring problem solves the partition problem.}
\begin{theorem}
\label{thm:exact-K-coloring-NPHard}
\red{Finding a weighted directed in-balanced $K$-coloring with fixed nodes is weakly NP-Hard.}
\end{theorem}

\begin{proof}
\red{Let $n$ and $X=\{x_1,\ldots,x_{n}$ be a given instance of partition. The graph we construct consists of $K-3$ nodes in a cycle and two in-stars. The first in-star has $3$ nodes and the other has $n+1$ nodes. Let $\omega$ be equal to the sum of the weights in $W$. The weights on the first in-star will all be identically $\omega/2$ and the weights on the second in-star will be $x_1,\ldots,x_{n}$. The weights on the edges in the cycle with $K-3$ nodes are $\omega, 2\omega, \ldots, (K-3)\omega$. If $K-3 = 1$, then add a loop on the node with weight $\omega$. More precisely, let $G = (S_1 \cup S_2 \cup C, E_1 \cup E_2 \cup E_C)$ where $S_1 = \{u_0,u_1,u_2\}$ and $S_2 = \{v_0,v_1,\ldots,v_{n}$. Let $C=\{c_1,\ldots,c_{K-3}\}$. Let $E_1 = \{u_1u_0,u_2u_0\}$ and $E_2 = \{v_iv_0:i=1,\ldots,n\}$. Let $E_C = \{c_ic_j: i < j\}$.  For $ij \in E_1 \cup E_2$, the weight on $ij$ is $w_{u_1u_0} = w_{u_2u_0} = \omega/2$ and
$w_{v_iv_0} = x_i$ for $i=1,\ldots,n$. For $ij \in E_C$, the weight is $w_{ij} = \omega$. Set $\cF = \{ \{u_0, v_0\}, \{u_1\}, \{u_2\} \}$. Note that this forces any in-balanced coloring to have $\omega/2$ weight going from nodes in a partition with $u_1$ to $v_0$ as well as the same $\omega/2$ weight going from nodes in a partition with $u_2$ to $v_0$. Note that if $K=3$, then both $C$ and $E_C$ are empty sets. If not then, then any balanced coloring must assign the nodes of $C$ distinct colors from all others, i.e., the nodes of $C$ are all isolated and use $K-3$ colors. Thus, the nodes $v_1,\ldots,v_n$ must be in a partition either with $u_1$ or $u_2$ as there are edges that enter $v_1,\ldots,v_n$. Thus, a partition exists if and only if there is a balanced coloring or corresponds exactly to the color $v_1,\ldots,v_n$ are assigned. }
\end{proof}

\red{We have an immediate Corollary.}
\begin{corollary}
\label{cor:directedORP-NPHard}
\red{The in-balanced ORP is weakly NP-Hard.}
\end{corollary}
\begin{proof}
\red{Note that if any instance of ORP is solved with a zero optimal objective, then the original graph and has a balanced $K$-coloring. }
\end{proof}

We are unaware of any other complexity results involving variants of balanced coloring/equitable partition. Interesting conjectures and open questions include the following.
\begin{itemize}
    \item We conjecture that finding an undirected weighted balanced coloring with fixed nodes is NP-Hard. 
    \item Famously, the complexity of graph isomorphism is unknown although many variants are efficiently solvable, including when the input graph has bounded degree~\cite{mckay14}. Given that finding a minimal balanced coloring is a preprocessing step for the graph isomorphism algorithms~\cite{mckay81,Darga04,Junttila07,Piperno08,Lopez09}, is there a connection between Theorem \ref{thm:exact-K-coloring-NPHard} and graph isomorphism?
\end{itemize}

\section{Linear Integer Program Formulation}
\label{sec:formulation}
In this section, we formulate Eq.~\eqref{eq:sumPseudo} as an integer linear program. We also note that our formulation extends to the directed case as well as to the addition of positive or negative weights. Modifying the formulation to account for edge perturbations   
is possible but would require the addition of a new family of variables. The latter changes could require the formulation to become a mixed-integer linear program although this would depend on whether the perturbation restriction was to a discrete set or not.

We consider a problem of form Eq.~\eqref{eq:sumPseudo}, i.e., we are given an undirected graph $G=(V,E)$ and a positive integer $K$ which denotes the number of colors we
require for our pseudobalanced coloring. We let $M = |V|-1$.

We use three sets of decision variables indexed over $V$, $E$, and $\cK
:= \{1,\ldots,K\}$. For every $p, q \in V$, $k \in \cK$, and $(i,j) \in
E^C$, let  
\[
  \begin{array}{ll}
    x_{pq} & = \left\{
             \begin{array}{ll}
               1 & \mbox{if nodes $p$ and $q$ are the same color}\\
               0 & \mbox{otherwise}
             \end{array} \right. \\[2mm] 
    y_{pk} & = \left\{
             \begin{array}{ll}
               1 & \mbox{if node $p$ is color $k$}\\
               0 & \mbox{otherwise}
             \end{array} \right. \\[2mm] 
    z_{ijk} & = \left\{
             \begin{array}{ll}
               1 & \mbox{if the pseudoedge of $(i,j)$ exists and $j$ is color
                                                 $k$}\\
               0 & \mbox{otherwise}
             \end{array} \right. \\[2mm]
  \end{array}
\]
Note that $z_{ijk}$ is an indicator of a pseudoedge that also models the
color of the node $j$. See Eq.~\eqref{eq:repair-edge-color} below for how the color assignment of $j$ controls this. Controlling the decision variable in this manner allows Eq.~\eqref{eq:approx-weight} to be linear. However, because each pseudoedge has both a forward and backwards direction, the sums of the variables over the possible color assignments must equal each other. See Eq.~\eqref{eq:repair-edge-equal} below for this constraint.

The natural objective function is to minimize the sum of pseudoedges. However, there are edges that we wish to incentify more than others, so we minimize the weighted sum of pseudoedges.
\begin{equation}
  \label{eq:minpseudoweights}
  \min \sum_{(i,j) \in E^C} c_{ij} \sum_{k \in \cK} z_{ijk}.
\end{equation}
Setting $c_{ij} = 1$ minimizes the sum of pseudoedges. For the graphs we solved, we found setting
\begin{equation}
\label{eq:randicweights}
c_{ij} = \frac{1}{d_i d_j},
\end{equation}
where $d_i$ is the degree of node $i$, for $(i,j) \in E^C$ produced the best results. We discuss this objective in more detail below.

We must ensure that every color is used at least once.
\begin{equation}
  \label{eq:everycolor}
  \sum_{i \in V} y_{ic} \geq 1, \quad k \in \cK.
\end{equation}

We must ensure that every node is assigned a color.
\begin{equation}
  \label{eq:onecolor}
  \sum_{k \in \cK} y_{ik} = 1, \quad i \in V.
\end{equation}

We must ensure that a color can not be assigned to two distinct nodes
unless they are indeed the same color.
\begin{equation}
  \label{eq:color-pair}
  x_{pq} + 1 \geq y_{pk} + y_{qk}, \quad p,q \in V, p \not= q, k \in \cK.
\end{equation}
Note that Eq.~\eqref{eq:color-pair} only prevents the same color from being
assigned to two nodes that are {\it different} colors. Nodes $p$ and $q$
with $x_{pq} = 1$ could be assigned different colors. Thus, to prevent this,
we also need the following constraints.
\begin{equation}
    \label{eq:color-pair-same}
    (1 - x_{pq}) \geq y_{pk} - y_{qk}, \quad (1 - x_{pq}) \geq y_{qk}-y_{pk},
    \quad p, q \in V, p \not= q, k \in \cK.
\end{equation}
We must enforce that pseudoweights are only assigned if the color is permitted.
\begin{equation}
  \label{eq:repair-edge-color}
  z_{ijk} \leq W y_{jk}, \quad (i,j) \in E^C, k \in \cK.
\end{equation}

We must ensure that the edge and pseudoedge weights from any pair of
nodes that are the same color agree to every other color.
\begin{equation}
  \label{eq:approx-weight}
  \begin{array}{ll}
      \left(\sum_{j\in V: pj \in E} A_{pj} y_{jk} +\sum_{j \in
      V:(p,j) \in E^C} z_{pjk}\right) -
      &\left(\sum_{j\in V: qj \in E} A_{qj} y_{jk} +\sum_{j \in
      V:(q,j) \in E^C} z_{qjk}\right) \\[2mm]
    & \leq M_{pq}(1-x_{pq}), \quad p,q \in V, k \in \cK, p\not=q.
  \end{array}
\end{equation}
Note that this inequality is not symmetric and must be posed twice for
every pair of nodes.

We must ensure that the sum of pseudoweights of the forward and backwards
edges agree.
\begin{equation}
  \label{eq:repair-edge-equal}
  \sum_{k \in \cK} z_{pqk} = \sum_{k \in \cK} z_{qpk}, \quad pq \in E'.
\end{equation}

In order to improve the formulation, we add the following antisymmetry constraints \cite{SheraliS01}.
\begin{equation}
    \label{eq:antisymmetry}
    \sum_{i \in V} y_{ik} \geq \sum_{i \in V} y_{i(k+1)}, \quad k=1,\ldots,K-1.
\end{equation}
This imposes an arbitrary ascending order on the size of the color sets.

The complete formulation for the instances we solve is then as follows.
\begin{equation}
\label{eq:mainip}
\tag{PBCIP}
  \begin{array}{lll}
     B^* := \min & \sum_{(i,j) \in E^C} c_{ij} \sum_{k \in \cK} z_{ijk} &
                                                   \eqref{eq:minpseudoweights}
    \\[2mm]
    \mbox{s.t.} & \eqref{eq:everycolor}, \eqref{eq:onecolor},
                  \eqref{eq:color-pair}, \eqref{eq:repair-edge-color}, 
                  \eqref{eq:approx-weight}, \eqref{eq:repair-edge-equal},
                  \eqref{eq:antisymmetry} & \\[2mm]
          & y_{ik}, x_{ij} \in \{0,1\} & i,j \in V, k \in \cK \\[2mm]
    & z_{ijk} \in \{0,1\} & (i,j) \in E^C, k \in \cK.                  
  \end{array}
\end{equation}

For some experimental samples, it is also possible to have advance knowledge about nodes that are of the same color. For instance, many biological networks studied in \cite{MoronePNAS20} contain clusters of nodes in perfect balanced coloring already. These perfect colors are found by the algorithm of Kamei and Cock \cite{Kamei13}. In these cases, it is important to keep these perfect colorings, and search for repairs that will not break these balanced  colors. In such a case, as a pre-processing step, for all nodes in perfect balanced coloring $u, v \in V$, we add the constraint:
\begin{equation}
    \label{eq:same-color}
    x_{uv} = 1.
\end{equation}

\section{Computational results for repairing a graph}
\label{sec:results}


Solving the integer linear program \eqref{eq:mainip} from Section \ref{sec:formulation} finds the minimum number of edges to add to a given graph to ensure a balanced coloring for a fixed number of colors. This is an important step towards our goal of finding pseudosymmetries of a graph. Our overall repair method is as follows and takes a candidate input graph with $n$ nodes.
\begin{enumerate}
    \item Pre-processing: Use the algorithm of Monteiro, {\it et al.}~\cite{Monteiro21} or Kamei and Cock \cite{Kamei13} to find a minimal balanced coloring present in the graph and identify the initial non-trivial colors. For the colored nodes, set Eq.~\eqref{eq:same-color}. Let $C$ denote the number of non-trivial colors and $T$ the number of trivial colors in this initial coloring.
    \item For $k=C$ to $C + T$ do the following:
    \begin{enumerate}
        \item Solve \eqref{eq:mainip} for $k$ colors, with constraints of form Eq.~\eqref{eq:same-color} added to fix the nodes that were already one of the $C$ non-trivial colors.
        \item Construct the resulting graph with the added edges dictated by \eqref{eq:mainip} and color the nodes using the algorithm of Monteiro, {\it et al.}~\cite{Monteiro21} again.
    \end{enumerate}
    \item {\it Evaluate} the resulting $T$ graphs to identify what is the number of colors of the {\it best} repaired graph employing {\it indices} characterizing graph topology and stability analysis, see below.
\end{enumerate}

In this section, we investigate different heuristics using graph topology and stability to accomplish step 3. We are aware of the methods presented in \cite{Morone19}, whose method uses pseudosymetries (Section \ref{sec:intro_and_motivation}) and applies repairs manually using expert knowledge, in \cite{Boldi21}, whose method uses quasifibrations  and applies repairs based on the similarity between nodes' input trees in directed graphs, in \cite{YanchenApprox20}, who devises a Monte-Carlo randomized method that minimizes Eq.~\eqref{eq:perm},  $\norm{[P, A]}$, and \cite{Gambuzza2020}, whose method applies repairs in order to create an idealized graph. In our approach we generate a series of potential solutions by repeatedly solving \eqref{eq:mainip} as described in steps 1 and 2. In order to accomplish step 3, a repaired graph is selected from this series of solutions by analyzing functions that characterize the underlying graph topology and stability of the solution. We call these graph functions {\it indices}.

\begin{figure}[ht]
    \centering
    \includegraphics[width=\linewidth]{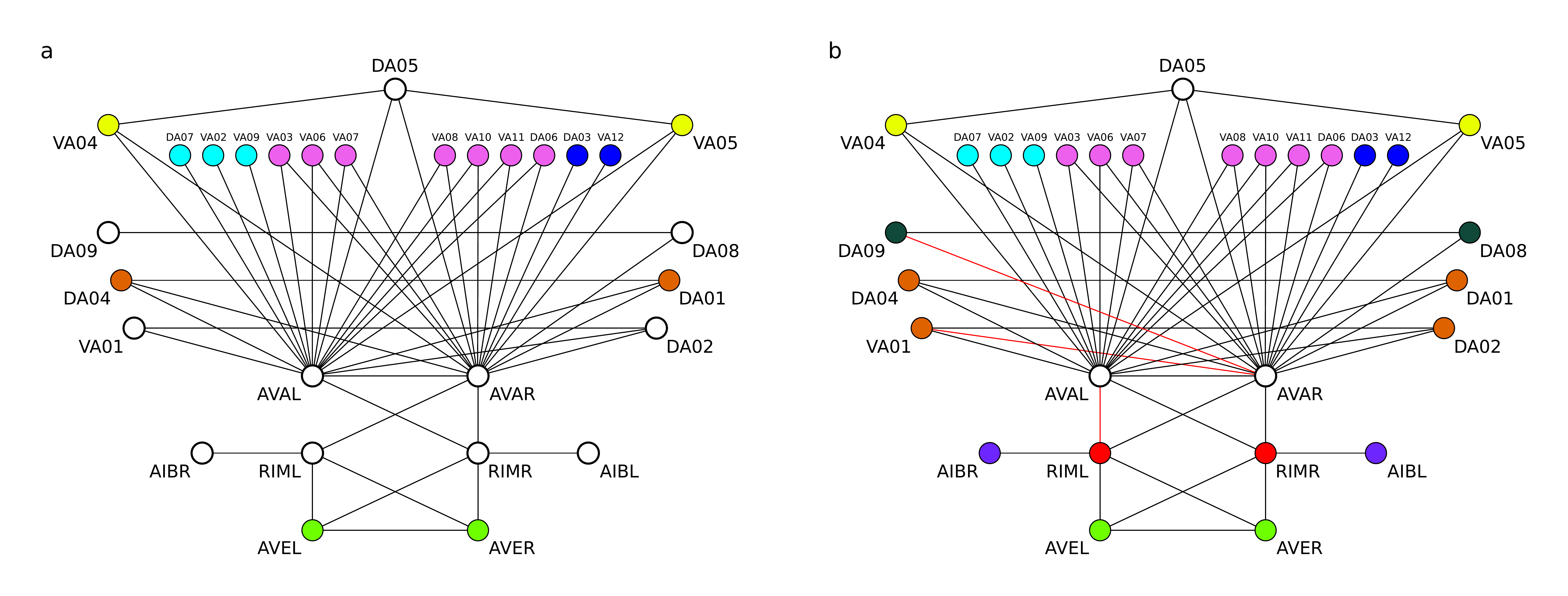}
    \caption{\textbf{(a)} The original backward circuit from \cite{Varshney11} with 17 colors and \textbf{(b)} the repaired circuit resulting from our method with 12 colors. White nodes in both represent nodes with trivial, i.e., individual unique colors.}
    \label{Fig:Backward_original_repaired}
\end{figure}


We consider the same networks analyzed in the pseudosymmetry analysis done by Morone {\it et al.} \cite{Morone19} from the connectome of {\it C. elegans} \cite{White86} obtained from the curated graph studied in \cite{Varshney11}. The graphs studied are induced  subgraphs of the connectome composed of neurons that are known to be involved in the locomotion function of the worm. These neurons have been identified in many experiments \cite{White86,Kato15} and are classified  into command interneurons (like AVBL, AVBR or AVAL, AVAR) and motor neurons (the series of neurons denoted by VA, DA, and VB, DB, for dorsal and ventral neurons).   The connectome is composed of two types of connections, gap junctions and chemical synapse. Around 10\% of the synapses are gap junctions. We test the algorithm only with undirected networks, so we use the connectome of gap-junctions connections between those forward and backward neurons. The connectome of chemical synapses is made of directed edges, and it is not used here, but in the test of the repaired algorithm of \cite{Boldi21} which uses quasifibrations and directed graphs. 
We  study two undirected gap junction networks referred to as the {\it forward circuit} and the {\it backward circuit}. These networks represent the neural circuitry responsible for forward and backward movement of the worm \textit{C. elegans}.

Following \cite{Morone19}, we obtained the data from \cite{Varshney11}. The undirected edges of this connectome are weighted since, generally, there are more than one gap junction connection between a pair of neurons. Typically, a neuron has a long axon which touches another neuron in different places along axon-to-axon or axon-to-soma contacts. The weight of the gap junctions is the number of those contacts. Here we will ignore the weight of the gap junctions and consider a binary adjacency matrix of 0 and 1 representing the absence or the present of a link.  The neural network of \textit{C. elegans} is known to have a $25\%$ variability between specimens \cite{White86, Varshney11}. 

Morone {\it et al.}  \cite{Morone19} presented a repair of these networks done manually by completing the network with the fewest possible links chosen by taking into account biological knowledge on the functions of the neurons. Although this manually repair network is not optimal, nor the true one, we will refer to this network as the 'ground truth' graph (or 'manual') with the ideal symmetry, so it provides an opportunity to test our more automated repair method.

\begin{figure}[ht]
    \centering
    \includegraphics[width=\linewidth]{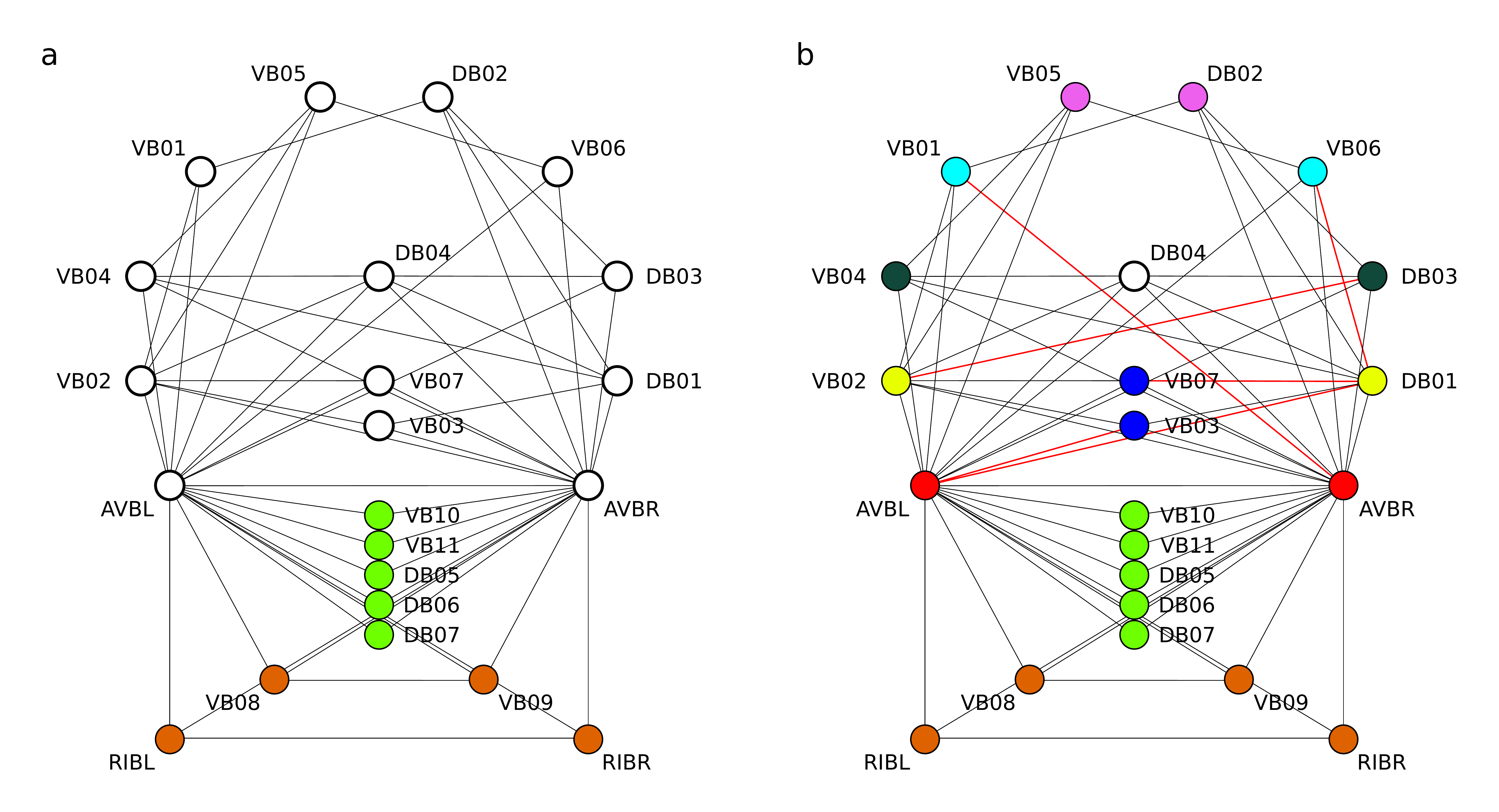}
    \caption{\textbf{(a)} The original forward circuit from \cite{Varshney11} with 15 colors and \textbf{(b)} the repaired circuit resulting from our method with 9 colors. White nodes in the original represent nodes with trivial, i.e., individual unique colors.}
    \label{Fig:Forward_original_repaired}
\end{figure}

The original (unrepaired) backward and forward circuits obtained from \cite{Varshney11} are shown in Figure~\ref{Fig:Backward_original_repaired}a and \ref{Fig:Forward_original_repaired}a, respectively. The original backward circuit contains 17 colors: 11 trivial and 6 non-trivial and forward circuit contains 15 colors: 13 trivial and 2 non-trivial. These colors are obtained using Kamei and Cock algorithm \cite{Kamei13} for perfect balanced coloring. Solving \eqref{eq:mainip} results in 12 possible repaired graphs for the backward circuit with the total number of colors ranging from 6 to 17 and 14 possible repaired graphs for the forward circuit with the total number of colors ranging from 2 to 15. In section \ref{section:step-by-step_repairs} we describe these graphs and in section \ref{section:indices} we justify our selection of the best repaired graphs with 12 colors and 9 colors for the backward and forward circuit, respectively. Our selected solution graphs are shown in Figures~\ref{Fig:Backward_original_repaired}b and \ref{Fig:Forward_original_repaired}b.

\subsection{Backward and forward circuit repairs}
\label{section:step-by-step_repairs}



\begin{figure}[ht]
	\centering
	\includegraphics[width=0.75\linewidth]{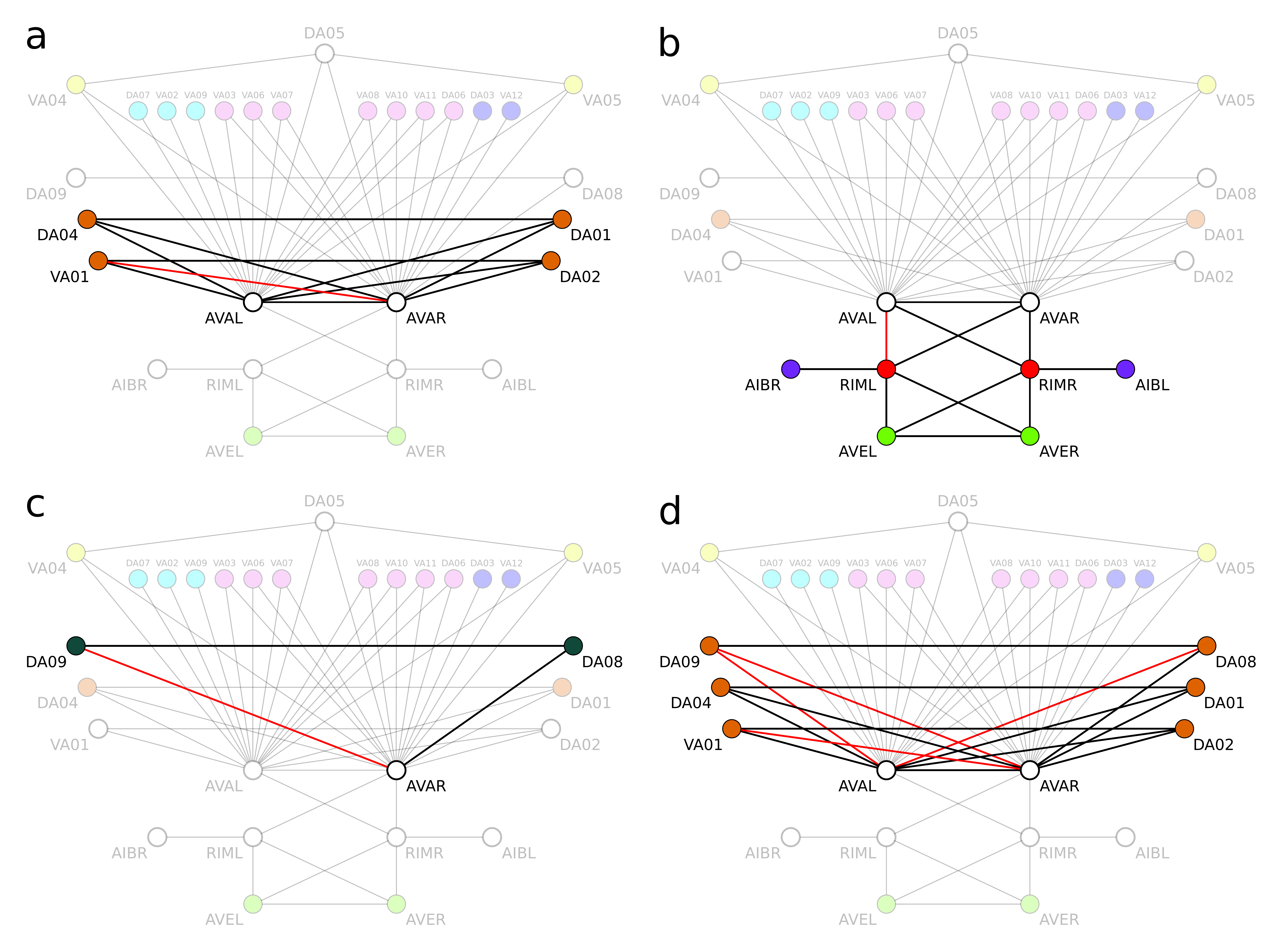}
	\caption{Repairs of the backward circuit I: \textbf{(a)} Edge (AVAR, VA01) colors nodes VA01 and DA02 orange together with DA01 and DA04, \textbf{(b)} Edge (AVAL, RIML) repairs the symmetry in the bottom part of the graph, \textbf{(c)} Edge (AVAR, DA09) puts nodes DA08 and DA09 in the same color, \textbf{(d)} Edges (AVAL, DA09), (AVAL, DA08), (AVAR, DA09), (AVAR, VA01) repair nodes VA01, DA01, DA02, DA04, DA08 and DA09 to the same color. This repair presents a slightly more complex version of repairs in (a) and (b).}
	\label{Fig:Backward_repairs_1}
\end{figure}

\begin{figure}[ht]
	\centering
	\includegraphics[width=0.9\linewidth]{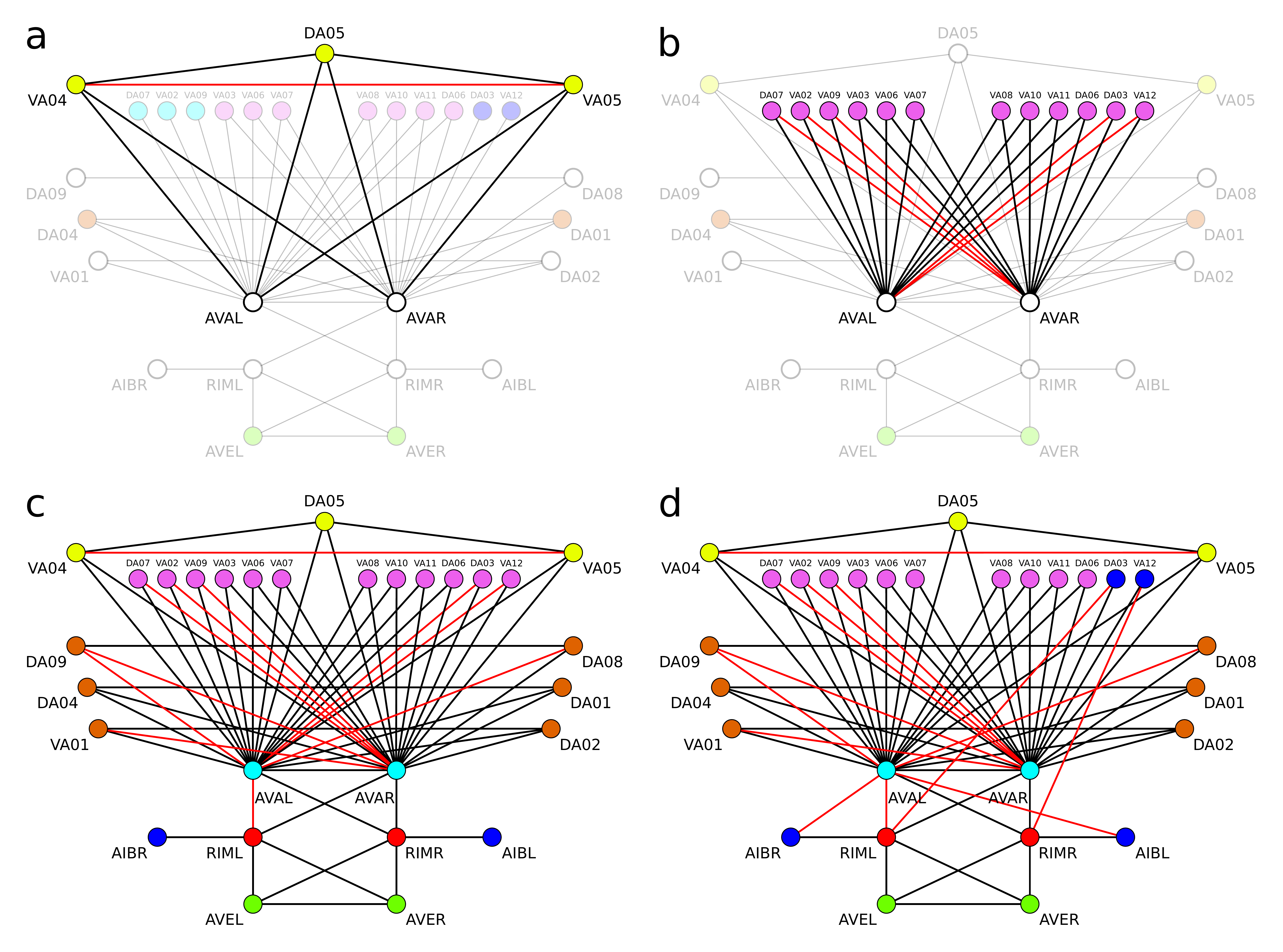}
	\caption{Repairs of the backward circuit II: \textbf{(a)} Edge (VA04, VA05) adds node DA05 to the yellow color. \textbf{(b)} (AVAL, DA03), (AVAL, VA12), (AVAR, DA07), (AVAR, VA02), (AVAR, VA09). \textbf{(c)} Repair of the backward circuit with 9 colors. \textbf{(d)} Repair of the backward circuit with 8 colors.}
	\label{Fig:Backward_repairs_2}
\end{figure}

We first describe the repairs of the backward circuit. We observe that the addition of just a few edges can drastically change the coloring. For example, adding the edge (AVAL, RIML) (as shown in Fig.~\ref{Fig:Backward_repairs_1}b) will make nodes RIML, RIMR, AIBR and AIBL symmetric. All repairs of the backward circuit can be decomposed into more simple repairs applied to the different part of the graph which are shown in figures \ref{Fig:Backward_repairs_1} and \ref{Fig:Backward_repairs_2}. Using this simple decomposition of repairs we demonstrate the way our method combines them to obtain compound repairs in Table~\ref{table:backward_repairs}. Note, repair Fig.~\ref{Fig:Backward_repairs_1}d is the combination of repair Fig.~\ref{Fig:Backward_repairs_1}a and Fig.~\ref{Fig:Backward_repairs_1}c, which are included in columns of Table~\ref{table:backward_repairs} corresponding to the number of colors between 11 and 9 for the simplicity of the interpretation. The repair with 9 colors shown in Fig.~\ref{Fig:Backward_repairs_2}c is a combination of all the simple repairs and all the repairs with 17-9 colors can be visualized using Figure~\ref{Fig:Backward_original_repaired}a and Table~\ref{table:backward_repairs}.

\begin{table}[ht]
    \centering
    \resizebox{\textwidth}{!}{
        \begin{tabular}{| c || c | c | c | c | c | c | c | c |} 
            \hline
            Repair & 17 Colors & 16 Colors & 15 Colors & 14 ( = 13) Colors & 12 Colors & 11 Colors & 10 Colors & 9 Colors \\
            \hline\hline
            Figure~\ref{Fig:Backward_repairs_1}a & & X & & X & X & X & X & X \\
            \hline
            Figure~\ref{Fig:Backward_repairs_1}b & & & X & X & X & X & X & X \\
            \hline
            Figure~\ref{Fig:Backward_repairs_1}c & & & & & X & X & X & X \\
            \hline
            Figure~\ref{Fig:Backward_repairs_1}d & & & & & & X & X & X \\
            \hline
            Figure~\ref{Fig:Backward_repairs_2}a & & & & & & & X & X \\
            \hline
            Figure~\ref{Fig:Backward_repairs_2}b & & & & & & & & X \\
            \hline
        \end{tabular}
    }
    \caption{Incidence matrix of edge repairs 1-6 from Fig.~\ref{Fig:Backward_repairs_1} and \ref{Fig:Backward_repairs_2} in the repairs of the graph corresponding to 17-9 colors. Each row represents the repair of the graph with different number of colors. As the number of colors decreases, the solution to \eqref{eq:mainip} utilizes more compounded combinations of repairs.}
    \label{table:backward_repairs}
\end{table}

Thus far, we have described the repairs with the number of colors between 9 and 17. The 8-color graph is shown in Fig.~\ref{Fig:Backward_repairs_2}d. This solution combines all the repairs above additionally symmetrizing nodes DA08 and DA09 with the cluster DA01, DA02, DA04 and VA01 and nodes AIBL and AIBR with nodes DA03 and VA12 in order to symmetrize hubs AVAL and AVAR. Repairs with less than 8 colors have a very complicated structure that is hard to visualize, and, as we will see later, the number of colors of the best solution lies far above this range, therefore we omit these repairs for simplicity.

Note that the 9 color repair in Fig.~\ref{Fig:Backward_repairs_2}c only has 7 colors. Recall, as described in step 2(a) of section \ref{sec:results}, the formulation \eqref{eq:mainip} has constraints of the form Eq.~\eqref{eq:same-color} to fix color of nodes that were known to be balanced as a result of preprocessing. Therefore, the colors of the fixed nodes DA07, VA02 and VA09 and nodes DA03 and VA12 cannot be changed or merged together with the pink color. To obtain the coloring in Fig.~\ref{Fig:Backward_repairs_2}c in step 2(b) our method obtains coloring using the minimal balanced coloring algorithm \cite{Monteiro21}, which combines all these nodes.

Another thing to notice is that the solutions with 14 and 13 colors are the same. This happens due to the fact that the solution obtained from solving \eqref{eq:mainip} with 14 colors assigns DA04 and DA01 a color that is different from the color of VA01 and DA02 even though the repair in Fig.~\ref{Fig:Backward_repairs_1}a is applied. In the 13 color solution obtained by solving \eqref{eq:mainip} all these nodes are assigned the same color as non-minimal balanced coloring is permitted and the minimum number of repaired edges required to obtain 13 colors is the same as that for 14 colors.


\begin{figure}[ht]
	\centering
	\includegraphics[width=0.8\linewidth]{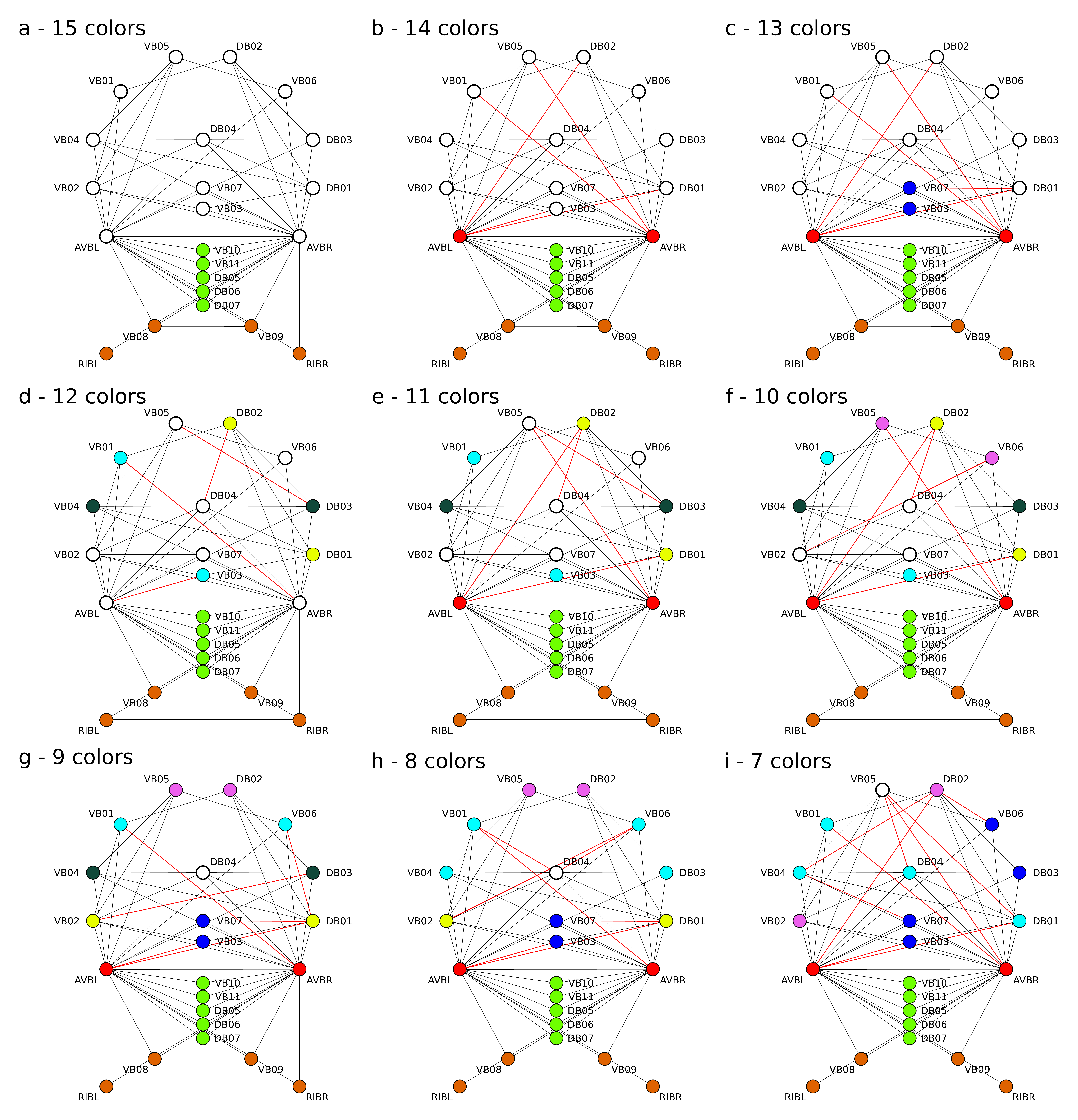}
	\caption{\textbf{(a)} The original forward circuit from \cite{Varshney11}, \textbf{(b-i)} Repaired versions of the forward circuit with 14-7 colors. Colors correspond to the minimal balanced coloring. Red links are repaired edges, i.e., edges not present in the original graph.}
	\label{Fig:Forward_repairs}
\end{figure}

Next, we present the repairs of the forward circuit. The topology of the forward circuit is quite different from the topology of the backward circuit. As we saw above, the backward circuit can be decomposed into a few independent parts that can be repaired separately. The forward circuit, on the other hand, can only be decomposed into two parts: the bottom part, namely nodes DB05, DB06, DB07, VB10, VB11, VB08, VB09, RIBL and RIBR that are symmetric and are therefore fixed by the formulation and the top part consisting of the rest of the nodes. The top part has high connectivity and the optimal repairs obtain solutions with symmetries that act on most of the nodes.
Instead of studying repairs as combinations as we did in the backward circuit, we examine all repairs separately. Figure ~\ref{Fig:Forward_repairs} shows the obtained repairs starting from the original circuit with 15 colors and ending with 7 colors. Table~\ref{table:forward_repairs} shows the incidence matrix of the edges used in these repairs. As before, we omit solutions with 2-6 colors since it will be shown that the best solution is not in this range. 

Observing solutions in Fig.~\ref{Fig:Forward_repairs} and Table~\ref{table:forward_repairs} we come to a few conclusions. First, we observe that due to the use of the objective that incentivizes the addition of edges between the high degree nodes, hubs are the first nodes to be repaired in the 14-color repair and they stay repaired almost throughout the rest of the repairs. Second, as the number of colors increases, the number of non-trivially colored nodes increases until almost none of the nodes are colored trivially. Third, the high connectivity of the top part of the circuit leads to instability in the use of the repaired edges and in the clustering of nodes. That is, edges that are used in the repairs with $K$ colors  are often not used in the repairs with $K - 1$ colors and nodes that have the same color for $K$ colors may not have the same color for $K-1$ colors.

\begin{table}[ht]
    \centering
    \resizebox{0.9\textwidth}{!}{
        \begin{tabular}{| c || c | c | c | c | c | c | c | c | c |}
            \hline
            Edge & 15 Colors & 14 Colors & 13 Colors & 12 Colors & 11 Colors & 10 Colors & 9 Colors & 8 Colors & 7 Colors \\ 
            \hline
            \hline
            VB02, VB06 &  & X &  &  & X &  &  &  &  \\ 
            \hline
            VB05, VB07 &  & X &  &  &  &  &  &  &  \\ 
            \hline
            AVBR, VB05 &  &  & X & X &  &  &  &  &  \\ 
            \hline
            DB01, VB05 &  &  & X &  &  &  &  &  & X \\ 
            \hline
            DB02, VB06 &  &  & X &  &  &  &  &  &  \\ 
            \hline
            VB03, VB05 &  &  &  & X &  & X &  &  &  \\ 
            \hline
            VB01, VB05 &  &  &  & X &  &  &  &  & X \\ 
            \hline
            DB03, VB04 &  &  &  & X &  &  &  &  &  \\ 
            \hline
            VB01, VB07 &  &  &  &  & X &  & X &  & X \\ 
            \hline
            DB03, VB02 &  &  &  &  & X &  &  & X &  \\ 
            \hline
            DB01, DB03 &  &  &  &  & X &  &  &  &  \\ 
            \hline
            AVBL, DB02 &  &  &  &  & X &  &  &  &  \\ 
            \hline
            AVBL, DB01 &  &  &  &  &  & X & X & X &  \\ 
            \hline
            DB04, VB06 &  &  &  &  &  & X &  & X & X \\ 
            \hline
            VB01, VB04 &  &  &  &  &  & X &  &  & X \\ 
            \hline
            DB02, VB02 &  &  &  &  &  & X &  &  &  \\ 
            \hline
        \end{tabular}
    }
    \caption{Incidence matrix of edge repairs in the Forward circuit shown in Fig.~\ref{Fig:Forward_repairs}. Each row represents a different edge and each column the repair of the graph with different number of colors. Some edges are omitted to improve the readability. The obtained solutions are inconsistent about which edges are repaired, i.e., the obtained solutions with $K - 1$ colors utilize edges that are very different from the ones used in the solution with $K$ colors. Note, some edges are omitted.}
    \label{table:forward_repairs}
\end{table}

\begin{figure}[ht]
	\centering
	\includegraphics[width=0.5\linewidth]{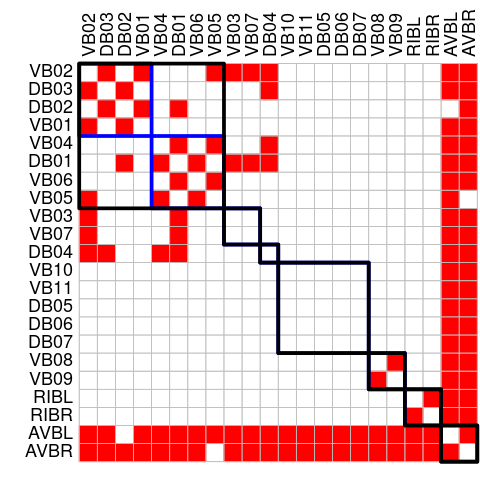}
	\caption{Circulant structure in the optimal repaired network found by the formulation in the forward circuit of Fig.~\ref{Fig:Forward_repairs}g and Fig. \ref{Fig:Forward_formulation_vs_nat_comm}a. This circulant structure is the same as found in the manually crafted solution in \cite{Morone19}. }
	\label{Fig:Fw_circulant_structure}
\end{figure}

Morone {\it et al.} \cite{Morone19} showed the importance of circulant structures in the locomotion networks. In particular, a circulant matrix in the adjacency matrix  corresponds to {\it cycles} in the network. A cycle is a path in a network plus an edge from the last node to the first, e.g., in Figure~\ref{Fig:Forward_original_repaired}b, VB02-DB03-DB02-VB01-VB02 or VB04-DB01-VB06-VB05-VB04. Cycles are important to generate  oscillatory behavior \cite{Morone19, Purcell10, LeiferPLOS20} and, as was mentioned before, the {\it C. elegans} locomotion process follows  oscillatory patterns. Therefore we look for circulant submatrices in the ideal solution for the forward circuit. In general, finding circulant submatrices in a given adjacency matrix is a problem on its own, and we are not aware of an algorithm that could do this automatically. For the small matrices considered here, we resort to find this circulant matrices by inspection, as done in \cite{Morone19}. Figure~\ref{Fig:Fw_circulant_structure} shows the adjacency matrix of the optimal forward circuit in Fig.~\ref{Fig:Forward_repairs}g and Fig. \ref{Fig:Forward_formulation_vs_nat_comm}a found by the formulation. The first eleven rows and columns represent the top part of the forward circuit (excluding hubs AVBL and AVBR) comprising most of the motor neurons. We note that the 8 by 8 submatrix in the upper-left corner of this adjacency matrix is (transpose) block circulant. That is, it is composed of two matrices. 
A circulant matrix that represents a 4-cycle permutation of VB02-DB03-DB02-VB01-VB02 and also of VB04-DB01-VB06-VB05-VB04:
\begin{equation}
  {\mathcal F} = {\rm
  circ}(0, 1, 0, 1) = \begin{bmatrix} 0 && 1 && 0 && 1\\ 1 && 0
    && 1 &&
    0\\ 0 && 1 && 0 && 1\\ 1 && 0 && 1 &
    &0\\
  \end{bmatrix},
\end{equation}
and a matrix ${\mathcal B}$ in the off-diagonal, such that the full 8 by 8 matrix is represented as:
\begin{equation}
\mathcal{BC}\ =   
\begin{bmatrix}
   {\mathcal F}  &  & {\mathcal B} \\
    {\mathcal B}^T &  & {\mathcal F}   
\end{bmatrix}.
\label{eq:bcirculant}
\end{equation} 
The transpose is needed due to the undirected character of the graph.
The same circulant structure, with exact the same loops represented by ${\mathcal F}$, has been found in the manually curated solution in \cite{Morone19}. This result is encouraging since it implies that the formulation is able to find not only close to optimal balance colorings but also by-product structures like circulant cycles in the network, which, in the particular case of locomotion, are necessary for the oscillatory motion of the animal.

\subsection{Indices on the graph}
\label{section:indices}

To identify the best repair over the number of $K$ colors, we consider a set of indices that characterize the graph topology and the stability of the dynamical solution imposed on the graph. We are looking for indices that have a qualitative change in their behavior that can indicate the most optimal solution of $K$ colors.

First, we list a few indices characterizing the partition induced by the balanced coloring. We consider the \textit{number of trivial colors} and the \textit{number of non-trivial colors} (NNTC) as a function of the $K$ colors. Each of these count the numbers of colors that are trivial and non-trivial, respectively. We also consider the \textit{number of nodes in non-trivial colors} which calculates the total number of nodes that belong to non-trivial colors.

\begin{figure}[ht!]
	\centering
	\includegraphics[width=0.4\linewidth]{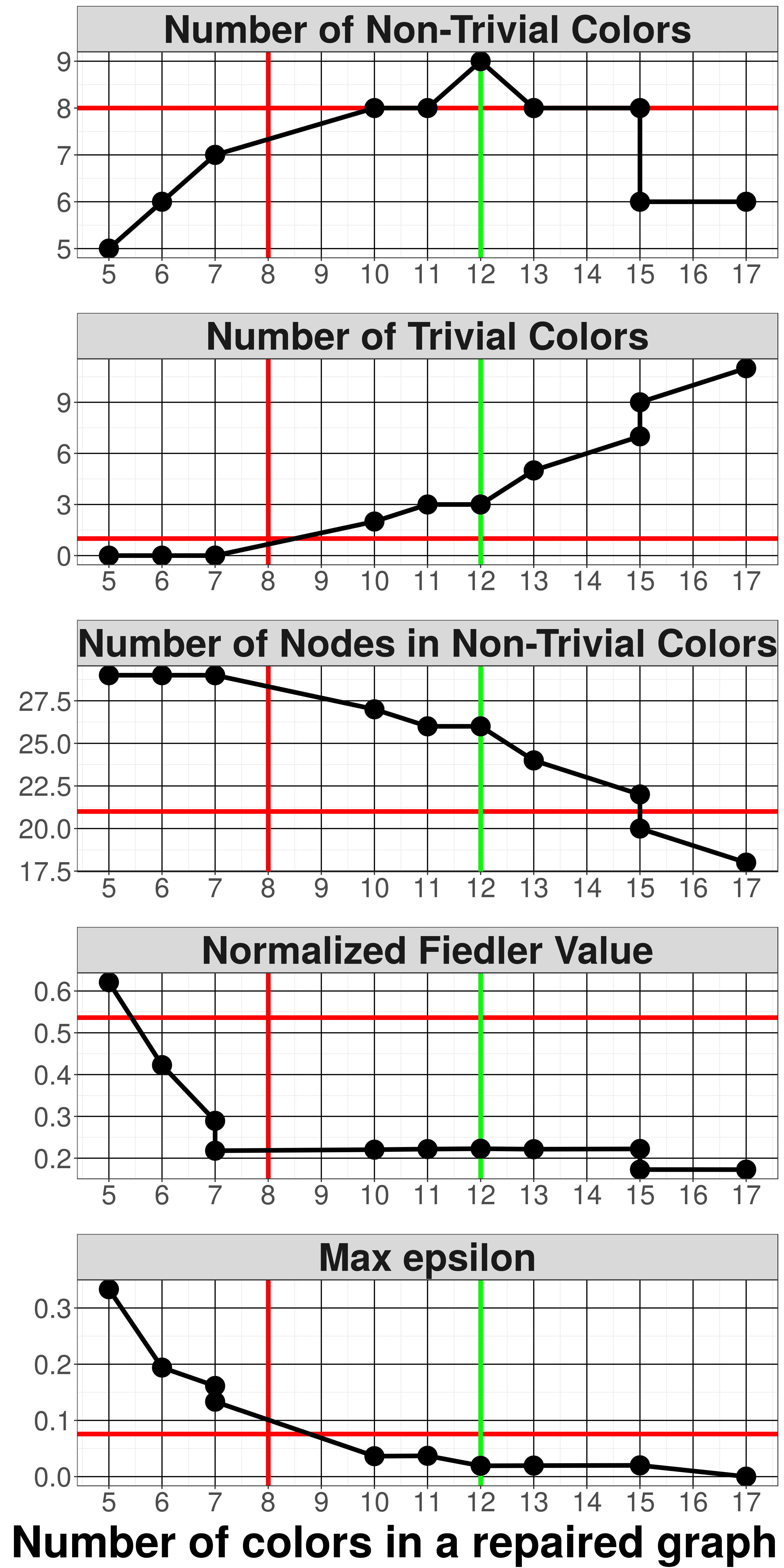}
	\includegraphics[width=0.4\linewidth]{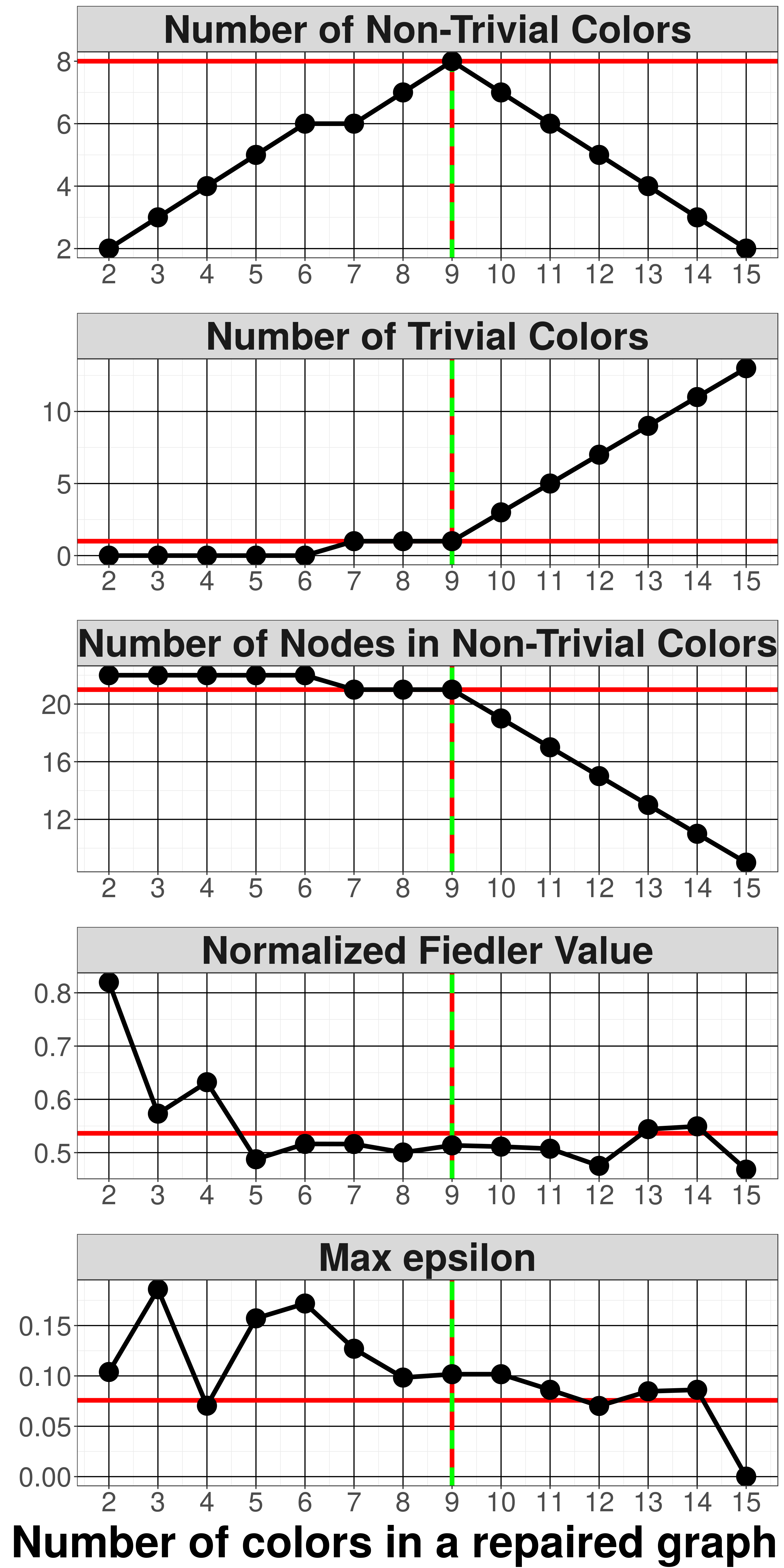}
	\put(-330,390){\Huge{\textbf{Backward}}}
	\put(-137,390){\Huge{\textbf{Forward}}}
	\caption{Color indices of the backward circuit (on the left) and forward circuit (on the right). Vertical red lines indicate the number of colors in a graph from Morone and Makse \cite{Morone19}. The horizontal red line indicates the value of the index for the same network. The green area on the left shows the optimal range of the number of colors for the backward circuit. The dashed green line on the right shows the optimal solution for the forward circuit.}
	\label{Fig:Indices}
\end{figure}

Figure \ref{Fig:Indices} (rows 1-3) shows the values of these indices obtained on the repaired graphs. First, we consider the number of non-trivial colors. As per the intuition stated in section \ref{sec:definitions}, a higher NNTC indicates the possible increase in the number of symmetries of the graph. Additionally, we can gain some insight by examining the limits of high number of colors and low number of colors. For a high number of colors, all non-fixed nodes will be in the separate classes, hence NNTC will be low. For a low number of colors a lot of non-fixed nodes will merge with fixed colors, thus NNTC will be low as well. We see in Fig.~\ref{Fig:Indices} that NNTC possesses two local minima for high and low numbers of colors and has a local maximum in between them. We use the solution corresponding to the local maximum as our best solution to pick the optimal number of colors which we call $K_{\rm opt}$. 

Using the NNTC index we  choose the best solutions for the forward and backward circuit to be at $K_{\rm opt}=9$ and 12 colors, respectively. The best solution is shown in green and the manual solution obtained in \cite{Morone19} is shown in red. Index values for the manual solution are shown with red horizontal lines. The number of colors for the best solution of the forward circuit was the same as the manual solution and the repairs used are similar. The number of colors for the backward circuit is different from the manual solution. We believe this is because the 12 color solution does not restore the symmetry between hubs creating more colors as shown by the dark and light blue colors in Fig.~\ref{Fig:Backward_original_repaired}b. We will further discuss these differences in Section \ref{sec:comparison_of_methods}. The number of trivial colors stays approximately constant below the best number of colors and starts increasing above it. The number of nodes in non-trivial colors behaves similarly, but decreases instead of increasing above the chosen solution.


The graph indices above characterize the coloring of the graph, but we cannot limit our consideration to the topology of the graph since biological networks and neural networks model systems that perform  functions vital for the organism's or ecosystem's survival. Therefore the dynamics of the network also needs to be accounted for. Next, we search for the optimal graph over $K$ by looking for the repair that could provide the largest stability to a given dynamical solution on the graph. In particular, we look for a graph that could provide a larger synchronizability, as defined by a larger stability of the cluster synchronization solution predicted by the balanced coloring on the graph.

Consider the eigenspectrum of the graph Laplacian. For a graph, $G=(V,E)$ with node-node binary adjacency matrix $A$, the random walk graph Laplacian \cite{BelykhDH05} is the matrix,
\begin{equation}
    \begin{aligned}
        L_G^{R} & = D^{-1}(D - A) = I - D^{-1}A = I - D^{-1/2} (D^{-1/2}A D^{-1/2})D^{1/2} \\
        & = D^{-1/2}(I - D^{-1/2}AD^{-1/2})D^{1/2} = D^{-1/2}L^{N}_G D^{1/2}
    \end{aligned}
\end{equation}
where $D$ is a $|V| \times |V|$ diagonal matrix with the $D_{ii}$ equaling the degree of node $i$ and $L^{N}_G$ is the normalized graph Laplacian. The matrix $L_G^R$  is symmetric, positive semidefinite, and singular so its eigenvalues are all nonnegative and real with at least one equaling zero. The eigenvalues of the graph Laplacian are well known to be associated with both the combinatorial and dynamical properties of the graph \cite{Fiedler73,Chung:1997fk,GhoshB06}. In particular we consider \textit{the normalized Fiedler value} defined as the second smallest eigenvalue of the random walk Laplacian. This has been shown to provide a measure of graph complete synchronizability \cite{wu1995application, zhou2006universality} for networks in which the response of the individual nodes is adjusted based on the number of incoming connections, which is consistent with the case of neural networks \cite{BelykhDH05}.  Complete synchronization of a graph is the state in which for all $i,j \in V : x_i(t) = x_j(t)$. Complete synchronizability, then, refers to the ability of the graph to achieve stable complete synchronization.

Different clusters of a biological network, e.g., clusters in the backward and forward circuits, perform different biological functions through cluster synchronization from the balanced colorings. These clusters are able to perform independent synchronized functions, and still be integrated in the network. Thus, a functional biological network requires cluster synchronization. Complete synchronization, instead, is deleterious for the organism. Therefore repairs that increase complete synchronizability of the graph are undesirable. 

Figure~\ref{Fig:Indices} (4th row) demonstrates the normalized Fiedler value for the obtained repairs. We observe that the normalized Fiedler value of the found best solution $K_{\rm opt}$ is in the range of the lowest values of all solutions. This behavior suggests that the best solution is one of the solutions that is less prone to the complete synchronizability, which is desirable for the dynamics of the system.

To summarize, we saw that the number of non-trivial colors can be used as an indicator for the best solution $K_{\rm opt}$. Therefore, in this paper we identify the best solution as the one with the highest NNTC. We also examined other indices such as the mean color size (the number of nodes divided by the number of colors), average clustering coefficient, average path length, the Randic index, the number of repaired edges and normalized Fiedler value, but they exhibited erratic behavior that did not seem indicative of any critical point on the studied graphs to choose the optimal one based on these indices. However, the normalized Fiedler value presents a plateau with minimum near $K_{\rm opt}$ (Figure~\ref{Fig:Indices}) indicating that there is a range of graphs with colors around $K_{\rm opt}$ that minimize the stability of the complete synchronization state which is beneficial for the biological network.

We believe that further work needs to be done in order to identify the best index, in particular, we believe the ideal index would need to more accurately characterize the synchronizability and stability of the dynamics on the graph. For instance, it would be desirable to analyze the stability of the cluster synchronization solution, rather than the instability of the complete synchronization solution as captured by the Fiedler value. 
The theory of cluster synchronizability is being developed in \cite{sorrentino21} and could provide a good index candidate to choose the optimal coloring solution. Here we only consider two networks that are fairly symmetric from the start and, in order to avoid overfitting, we choose NNTC as the graph function to select our solutions. An interesting open question is then to find the best index or indices to use to select the best candidate solution.


\section{Automorphism groups and pseudosymmetries of the repaired graphs}
\label{sec:normal_subgroup_decomposition}

In this section, we describe how the factorization into normal subgroups  of the automorphism group of a graph \cite{macarthur08} performed in the {\it C. elegans} connectome in \cite{Morone19}  partitions the graph into functional clusters, and how this partition can be obtained from \eqref{eq:mainip}.   We emphasize that a repair of the graph guided by balanced coloring provides a less stringent condition on the number of repaired edges than a repair following the restoration of the full symmetry automorphism group as done in \cite{Morone19}. That is, repairing the network to achieve perfect balanced coloring will always require less (or at most equal) number of edges than repairing the network with automorphisms. This is because, automorphisms impose more strict conditions on the structure of the network than balanced colorings and fibrations. This is particularly true when the graph is directed and one is only interested in cluster synchronization. Cluster synchronization is exclusively determined by the information that a node receives through its in-degree from the entire network. This information is taken into account either by the input tree of the node as in the fibration formalism of \cite{MoronePNAS20} or by the analogous  in-balanced coloring as considered here. The out-balanced coloring is irrelevant for cluster synchronization (see \cite{MoronePNAS20} for further details). 

Instead, the symmetries imposed by automorphisms require invariance of the full adjacency, including the in and out-degree. Thus, automorphisms require more stringent conditions than what is required to achieve cluster synchronization. Nevertheless, Ref. \cite{Morone19} showed that automorphisms exist in the connectome, in particular in the undirected gap junction connectome where the distinction between in and out balanced is meaningless.  These symmetries pose the particular property of factorization of the symmetry group, and this factorization separates the neurons into known classes like interneurons, motor and touch neurons. Thus, this particular symmetry group has a functional connotation. Therefore, next, we relate the pseudobalanced coloring formalism with the pseudosymmetry formalism of \cite{Morone19}.

For a graph, $G=(V,E)$, let $\Aut(G)$ denote the automorphism group of $G$. For any permutation $p \in \Aut(G)$, the {\it support of $p$} \cite{mckay81, macarthur08, Morone19}, denoted by $\supp(p)$ is defined as the nodes which do not have the same labels after the permutation $p$ has operated on $G$. Formally, this can be written
\begin{equation}
    \supp(p) := \{i \in V \mid p(i) \neq i\}.
\end{equation}
Two permutations, $p, q \in \Aut(G)$, are said to be {\it support-disjoint} if the node labels they change are completely different, i.e., if $\supp(p) \cap \supp(q) = \emptyset$. Two sets of permutations $P, Q \subseteq \Aut(G)$, are support-disjoint if every permutation in $P$ is support disjoint with every permutation in $Q$.  Note that any two support-disjoint permutations commute.

Let $S$ denote the set of generators of an automorphism group $\Aut(G)$. Partition $S$ into $n$ support-disjoint subsets $S = S_1 \cup S_2 \cup \dots \cup S_n$ such that none of the $S_i$ can be further decomposed into smaller support-disjoint subsets. Each $S_i$ generates a group $H_i$ that is a subgroup of $\Aut(G)$. Since all $S_i$ are support-disjoint, $H_i$ commute with each other and, since $S$ is a union of all $S_i$, $\Aut(G)$ is a direct product of all $H_i$:
\begin{equation}
    \Aut(G) = H_1 \times H_2 \times \dots \times H_n.
\end{equation}

All $H_i$ are normal subgroups, since they commute with the rest of the group. Therefore, partitioning the set of generators of an automorphism group into irreducible support-disjoint subsets creates a factorization of this group. The uniqueness and irreducibility of this representation has been shown in \cite{macarthur08}. Note, $H_i$ acts on the separate non-intersecting subsets of nodes defining a partition (also referred to as a factorization) on the nodes of a graph. We call the equivalence classes of nodes created by this partition \textit{sectors}~\cite{Morone19}.

Therefore, we have a way to factorize the graph via the automorphism group associated with it. We describe how to classify the obtained factors. Each factor is vertex-transitive, but vertex-transitive graphs are too broad a class to admit a complete classification. Nevertheless, we apply a simple computational approach that allows us to classify most of the obtained subgroups. In order to classify each normal subgroup we determine whether it is isomorphic to a symmetry group from this list: a symmetric group, a dihedral group, a cyclic permutation group, or an alternating group of sizes 1 to $n$, where $n$ is the number of nodes in the graph. If the normal subgroup is isomorphic to one of these classes, then we assign this class to it. The implementation uses NAUTY \cite{Mckay07}, Sympy \cite{meurer17} and Sage \cite{sagemath} and is available at \url{https://github.com/makselab/PseudoBalancedColoring}.



To illustrate the decomposition process, we apply the algorithm to the repaired backward circuit. The automorphism group is decomposed into

\begin{equation}
    \Aut(G) = H_1 \times H_2 \times H_3 \times H_4 \times H_5 \times H_6 \times H_7 \times H_8,
\end{equation}
where
$H_1 = D_4$ applied to nodes DA01, DA04, DA02 and VA01,
$H_2 = S_2$ applied to nodes VA04 and VA05,
$H_3 = S_2$ applied to nodes AVEL and AVER,
$H_4 = S_7$ applied to nodes DA06, VA03, VA06, VA07, VA08, VA10 and VA11,
$H_5 = S_2$ applied to nodes DA03 and VA12,
$H_6 = S_3$ applied to nodes DA07, VA02 and VA09,
$H_7 = S_2$ applied to nodes AIBR, AIBL, RIML and RIMR,
$H_8 = S_2$ applied to nodes DA08 and DA09.

\begin{figure}[ht]
	\centering
	\includegraphics[width=0.9\linewidth]{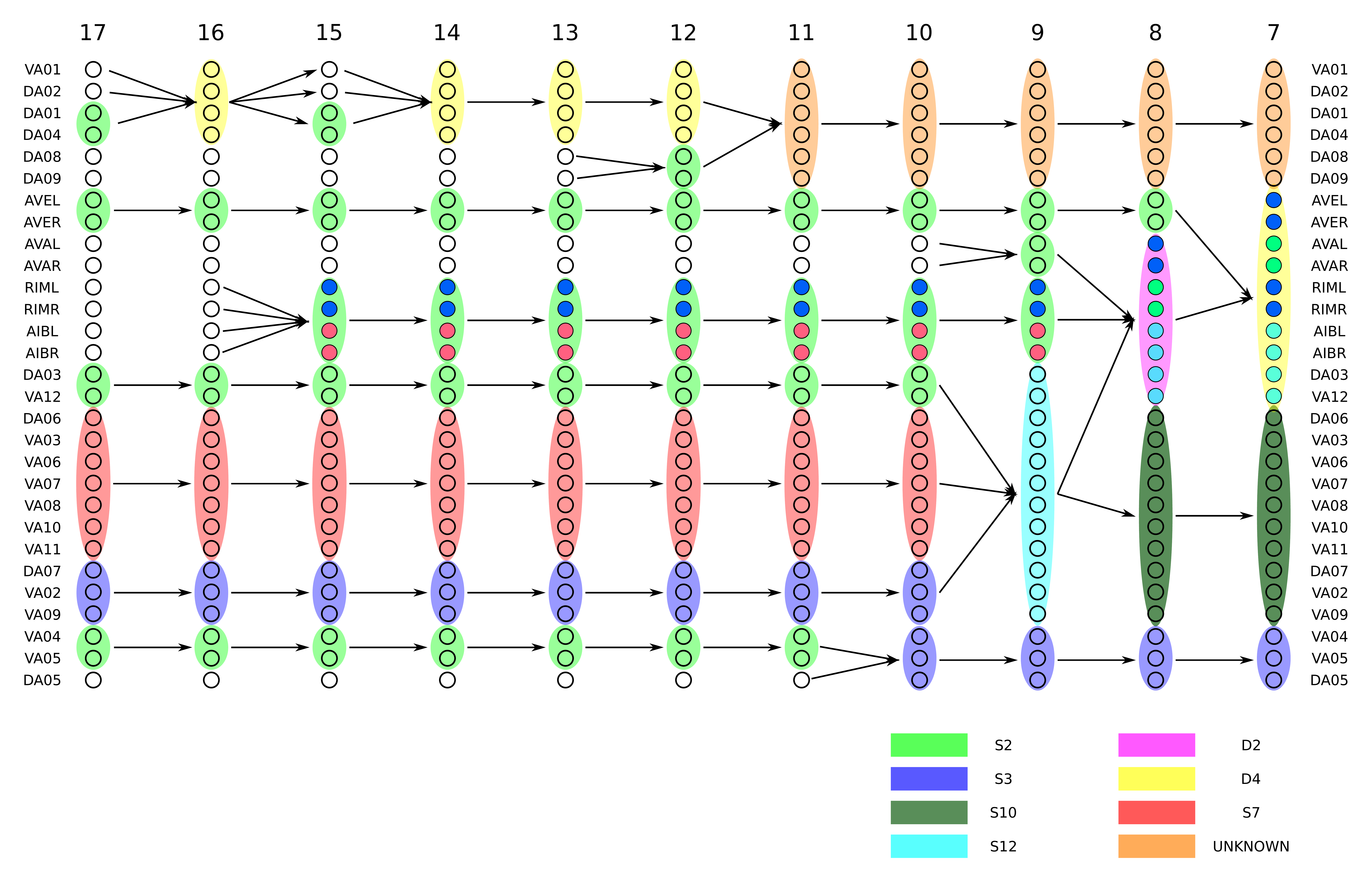}
	\caption{Step-by-step transformation of the backward circuit automorphism group. Each column corresponds to the repair with $K$ colors. Each row corresponds to one node. Colored circles show 7 classes of the normal subgroups. Nodes are colored inside the sector according to the orbit they belong to. For example, group $S_7$ in 15 color repair has only one orbit, hence all the nodes are of the same color. Group of RIMR, RIML, AIBL, AIBR in 15 color repair has two orbits which are shown with red and blue colors.}
	\label{Fig:Backward_aut_group}
\end{figure}

We apply this algorithm to the repaired graphs obtained by our algorithm on the backward and forward circuits. The result of this analysis for the backward circuit is shown in Fig.~\ref{Fig:Backward_aut_group}. Similar to the earlier presented conclusions from Table~\ref{table:backward_repairs}, we see here that between 17 and 9 colors, the formulation \eqref{eq:mainip} combines different simple repairs to obtain a symmetrized version of the network. This is evident by the fact that almost all new subgroups for $K - 1$ colors are created by merging some nodes (or subgroups) in $K$ colors and nodes never move between subgroups.
Note, our computational approach was able to classify all of the subgroups aside from the subgroup marked as "UNKNOWN" in the backward circuit.

\begin{figure}[ht]
	\centering
	\includegraphics[width=0.9\linewidth]{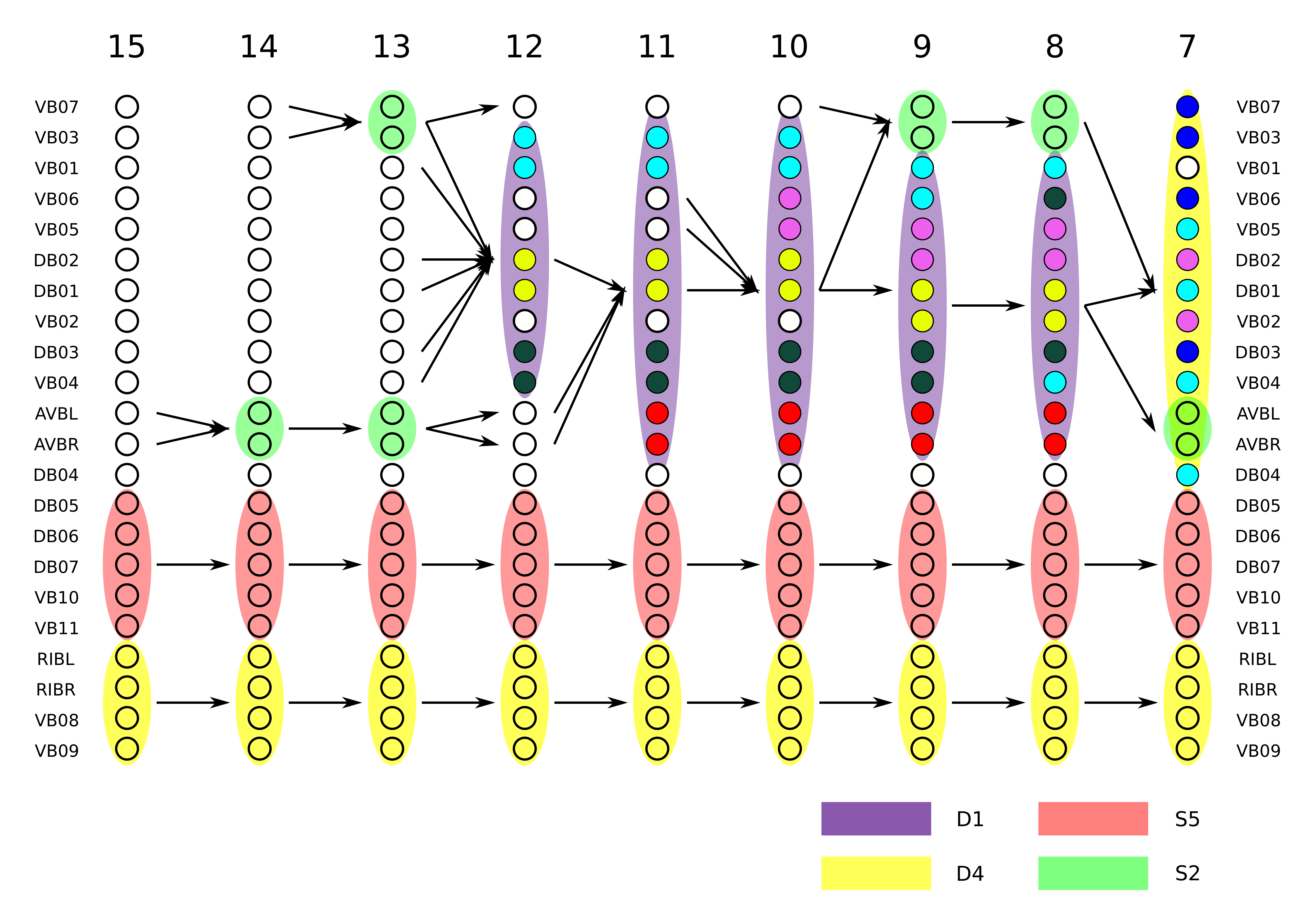}
	\caption{Step-by-step transformation of the forward circuit automorphism group. Each column corresponds to the repair with $K$ colors. Each row corresponds to one node. Colored circles show 3 classes of the normal subgroups. Nodes are colored inside the sector according to the orbit they belong to. Colors of orbits correspond to node colors in Fig.~\ref{Fig:Forward_repairs}. Nodes of the white color inside the sector don't belong to the sector. For example, group with nodes DB03, DB04, VB05 and VB06 in 13 color repair contains two orbits shown with yellow and blue colors and nodes DB02, VB01 and VB04 don't belong to this sector.}
	\label{Fig:Forward_aut_group}
\end{figure}

The results for the forward circuit are shown in Fig.~\ref{Fig:Forward_aut_group}. As observed before, the forward circuit is decomposed into two parts: a top part and a bottom part. As expected, the bottom part of the graph represented by the bottom 9 nodes of Fig.~\ref{Fig:Forward_aut_group} stays unchanged. The top part, as seen before, does not have a consistent partition and an increase in the number of colors combines different nodes in normal subgroups. Two things can be observed: 1) as the number of colors decreases, the number of symmetric nodes increases, and 2) nodes are mostly combined to exhibit mirror symmetry ($D_1$). Note that groups $D_1$ and $S_2$ are isomorphic and we use them interchangeably, $D_1$ is more appropriate in a case of mirror symmetry and is therefore used here. For example, in Fig.~\ref{Fig:Forward_repairs}g there is mirror symmetry in the top part between the nodes on the left and the nodes on the right.


To complete our analysis of the repaired networks, we find $\varepsilon$ described in Section \ref{sec:intro_and_motivation}. $\varepsilon$ represents the cutoff of the norm in the equation \eqref{Eq:psedosymm_natcomm} i.e.
\begin{equation}
    \norm{[P_\varepsilon,A]} \leq \varepsilon
\end{equation}
where $P_\varepsilon$ represent all permutations in automorphism group of the repair of a graph $G$. We denote the repaired graph $G_\varepsilon$; $\varepsilon$ then can be found using $\Aut(G_\varepsilon)$ as:
\begin{equation}
    \varepsilon = \max_{P_\varepsilon \in \Aut(G_\varepsilon)} \norm{[P_\varepsilon,A]}.
    \label{eq:maxEpsilon}
\end{equation}

Due to the size of the automorphism group, finding $\varepsilon$ using the entire automorphism group is computationally intractable. To give an approximation, we find $\varepsilon$ on the set of generators of the normal subgroups. That is, we first calculate the maximum $\varepsilon$ on generators of each normal subgroup and then we take the resulting $\varepsilon$ as their maximum. To simplify the interpretation of the obtained $\varepsilon$, we define
\begin{equation}
    \epsilon = \frac{\varepsilon}{4M},
\end{equation}
where $M$ is the total number of edges in the amended graph and 4 is the constant appearing due to the circumstances described in Section \ref{sec:intro_and_motivation}. Then, $\epsilon$ is the maximum of the number of edges that needs to be added to the original graph in order to make the permutation $P_\varepsilon$ a part of the automorphism group of the amended graph normalized by the total number of edges in the graph. Figure \ref{Fig:Indices} (last row) shows the values of $\epsilon$ for the obtained repairs. We see that all obtained repair graphs are well below 0.25 ($25\%$, except the backward repair with 5 colors) which is the value found in \cite{Varshney11} to be associated with the natural variability in the number of links that are different across the five animals that have been used to build the connectome in \cite{White86}. Therefore, from the point of view of pseudosymmetries, each of the repaired networks can be a candidate for the ``blueprint'' ideal symmetry network of {\it C. elegans}. Note that Fig.~\ref{Fig:Indices} shows the maximum values of $\epsilon$ as defined in Eq.~\eqref{eq:maxEpsilon}.

\section{Comparison with other link prediction methods}
\label{sec:comparison_of_methods}

\begin{figure}[ht]
	\centering
	\includegraphics[width=0.9\linewidth]{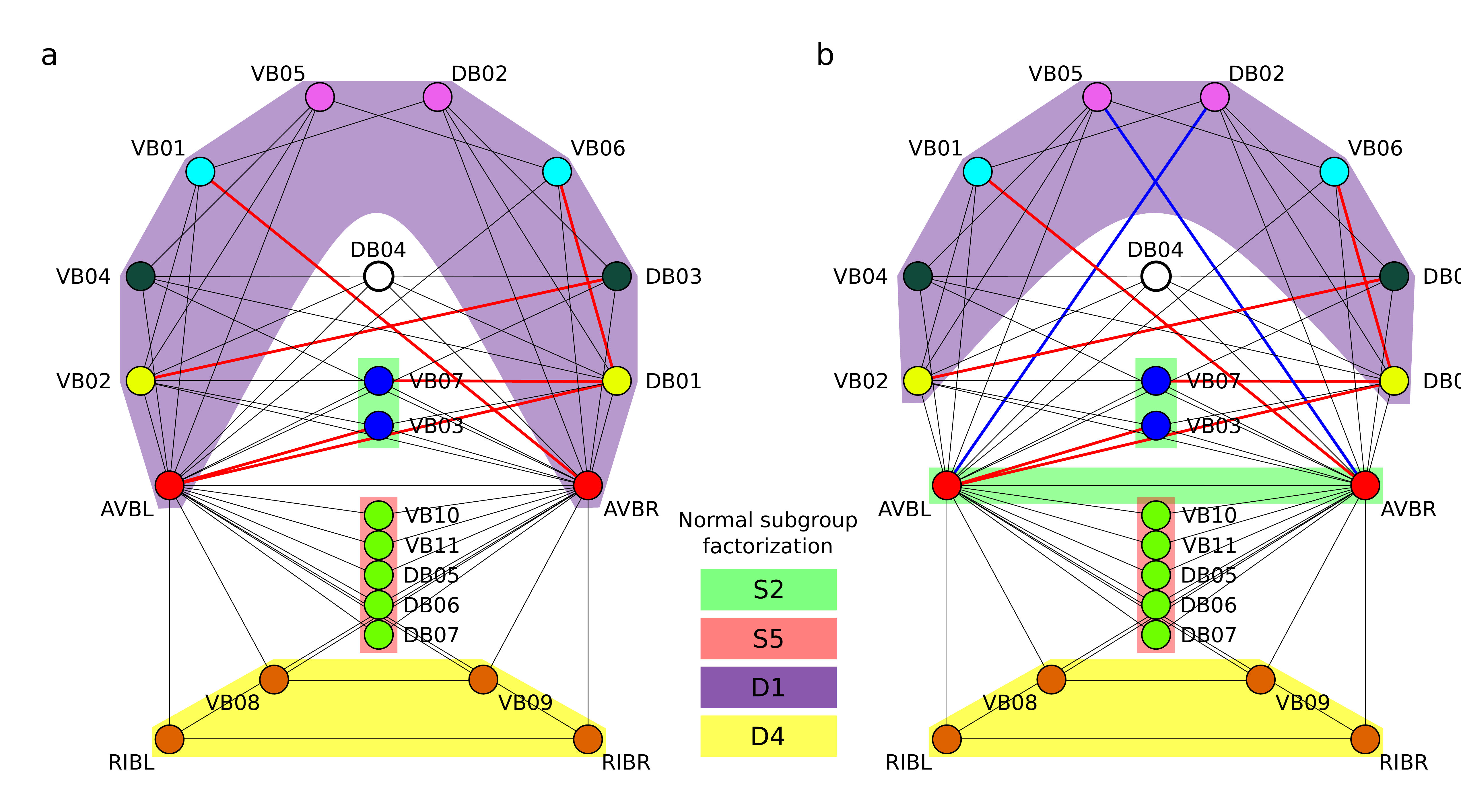}
	\caption{Comparison between the most optimal repair solution found by the formulation  \textbf{(a)} and manually crafted 'ground truth' solution \textbf{(b)} from \cite{Morone19} for the forward circuit. Red edges are repaired in both solutions and blue edges are repaired only in the manual solution. Background colors show the sectors associated with the normal subgroup factorization. Notice the complete agreement between both repairs regarding the colors, and almost complete agreement regarding the pseudoedges and normal subgroup factorization. The pseudo edges are the same except for two extra blue edges in the manual solution. This does not affect the colors. AVBL-AVBR belongs to the motor group $D_1$ in the formulation but they are an independent sector and subgroup $S_2$ in the manual solution. The blue edges were added to the manual solution in \cite{Morone19} to factorize the command interneurons AVB from the motor sector, but the formulation found another more optimal solution.  }
	\label{Fig:Forward_formulation_vs_nat_comm}
\end{figure}

To complete our analysis, we compare the results obtained by the formulation \eqref{eq:mainip} against the results of the manual repairs obtained in \cite{Morone19} and some of the traditional link prediction methods.

We begin by comparing our results with the manual repair 'ground truth' of \cite{Morone19}. Figure ~\ref{Fig:Forward_formulation_vs_nat_comm} shows the ideal (a) and manual (b) solutions for the forward circuit. The bottom part was left unchanged in both solutions. The top part of the circuit is very similar between solutions: coloring is exactly the same and the only difference is the addition of edges (AVBL, DB02) and (AVBR, VB05) in the manual solution. The similarity between these solutions can be ascribed to the fact that both repair methods have a similar objective. Morone {\it et al.} ~\cite{Morone19} obtained their manual solution by repairing the symmetry between the hubs and adding edges to complete the symmetry between the rest of the nodes. Likewise, the formulation \eqref{eq:mainip} with the weighted objective in Eq.~\eqref{eq:randicweights} incentifies repairs between nodes with higher degrees (hubs). 

The difference between the solutions can be interpreted by observing that the manual solution decomposes the automorphism group of the top part into $S_2$ (AVBL, AVBR) $\times \, D_1$ (VB02, VB04, VB01, VB05, DB02, VB06, DB03, DB01) $\times \, S_2$ (VB03, VB07), while our solution decomposes it into $D_1$ (AVBL, AVBR, VB02, VB04, VB01, VB05, DB02, VB06, DB03, DB01) $\times \, S_2$ (VB03, VB07). Colorings corresponding to these decompositions are the same and therefore, because the objective of \eqref{eq:mainip} is to minimize the weighted sum of edges added,  adding two extra edges is not optimal. However, the logical assumption that the function of the hubs is different from the rest of the nodes led \cite{Morone19} to dividing hubs into a separate subgroup. We conjecture that the solutions to \eqref{eq:mainip} would have separated hubs in a bigger network in order to symmetrize other parts of the network connected to them.

\begin{figure}[ht]
	\centering
	\includegraphics[width=\linewidth]{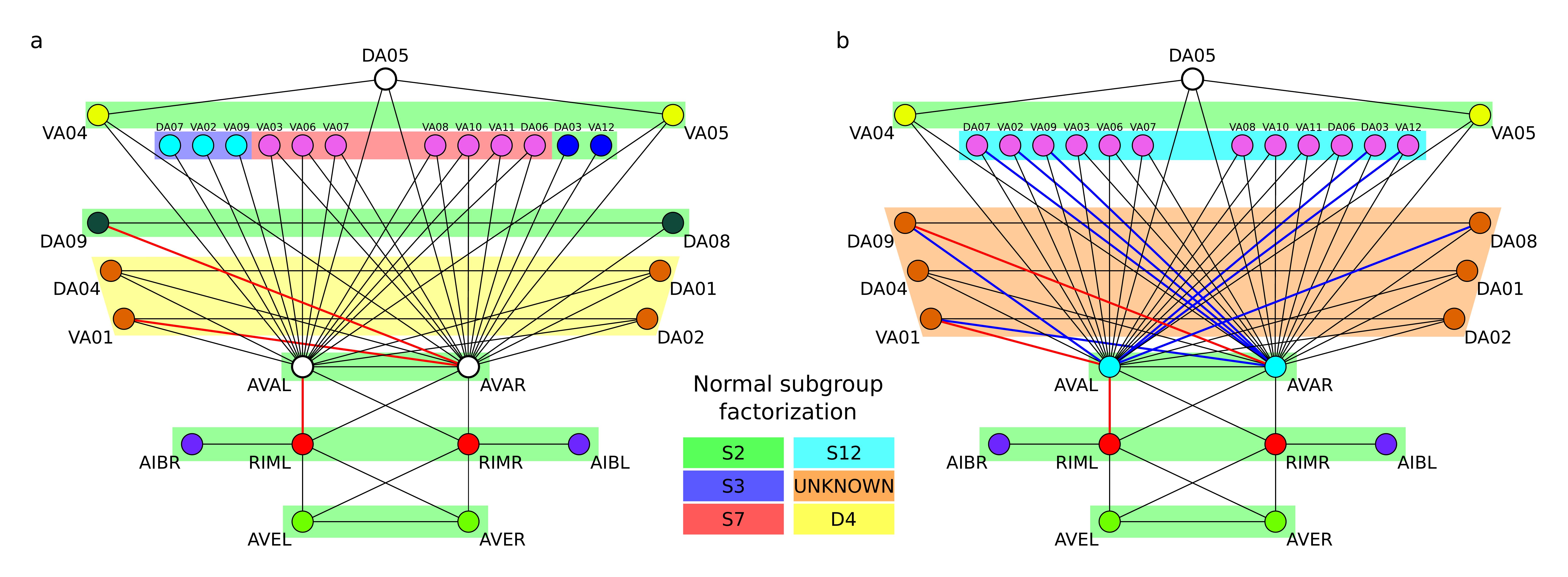}
	\caption{Comparison between the most optimal repair solution found by the formulation  \textbf{(a)} and the manually crafted 'ground truth' solution \textbf{(b)} from \cite{Morone19} for the backward circuit. Red edges are repaired in both solutions and blue edges are repaired only in the manual solution. Background colors show the sectors associated with the normal subgroup factorization. The crucial AVAL-RIML link is found by both solutions. However the formulation finds that the $S_2$ normal subgroup of the manual solution can be repaired with fewer pseudoedges by breaking this subgroup in three sectors as shown. As in the  forward circuit, Fig. \ref{Fig:Forward_formulation_vs_nat_comm}b, the manual solution attempts to factorize the interneurons AVAL-AVAR from the motor sector, at the expense of creating more pseudoedges than needed. In contrast, the formulation finds a better solution with less pseudoedges. }
	\label{Fig:Backward_formulation_vs_nat_comm}
\end{figure}

Figure ~\ref{Fig:Backward_formulation_vs_nat_comm}a shows the ideal solution obtained by our formulation and Fig.~\ref{Fig:Backward_formulation_vs_nat_comm}b shows the manual solution for the backward circuit. We can separately compare the repairs applied to different parts of the circuit. Firstly, repairs (AVAL, RIML) and (VA01, AVAR) were obtained in both solutions. Secondly, nodes DA08 and DA09 were repaired to be symmetric similar to the manual version, but links between these two nodes and AVAL needed to symmetrize them with the rest of orange nodes weren't repaired. Lastly, hubs AVAL and AVAR along with the manually restored nodes with $S_{12}$ symmetry in pink were not repaired. The solution with 9 colors in Fig. \ref{Fig:Backward_repairs_2}c does symmetrize AVAL-AVAR, all of the pink nodes and DA08-DA09, but creates an extra symmetry between nodes DA05, VA04 and VB05. Hence, this repair can be obtained, but it wasn't chosen due to the fact that the maximum of NNTC corresponds to the solution with 12 colors. This provides evidence that further analysis of indices is needed to improve the way to identify the best solution.


Now we take the manual repair as the ground truth and compare the performance of the formulation with some of the traditional link prediction methods. The link prediction problem is a problem of predicting the probability of the existence of the edge between two not connected nodes given the observed data. This probability is defined as the similarity between considered nodes obtained using a topological measure of the graph \cite{Newman08, Liben07, Lu11}. To repair a graph using a link prediction approach, an empirically chosen number of top ranked edges is predicted to exist \cite{Chen05}. Commonly used measures of similarity are divided into local, global and quasi-local classes. Further, we will define some of them following \cite{Lu11} and propose a way to estimate their performance.

First, we introduce a few indices characterizing the local topology of the graph. Let $S_{xy}$ be the distance between nodes $x$ and $y$, $\Gamma(x)$ be the neighborhood of node $x$ (set of nodes connected to $x$), $\mid X \mid$ be the cardinality of set $X$ and $k_x$ be the degree of node $x$. \textit{Preferential attachment index} calculates the similarity between two nodes based on their degree:
\begin{equation}
    S_{xy}^{PA} = k_x \times k_y.
\end{equation}

This measure is widely used in scale-free networks and is generalized by the Randic index~\cite{randic75, KP10}. In particular, $s(G) = \sum_{x, y \in V} S_{xy}$ is called a scale-free metric and is used as a measure of scale-freeness of the graph \cite{Li05}. A degree based metric is also used while generating a random scale-free network with the Barabasi-Albert model \cite{Albert02}. In this model, preferential attachment implies that each new added node in a generated graph is connected to other nodes with the probability proportionate to their degree.

Common Neighbors similarity metric is based on the assumption that nodes that have a lot of common neighbors are likely to be neighbors themselves. This assumption has had success in social networks \cite{Newman01, Kossinets06}, however, as will be discussed in more detail later, it does not increase the symmetry of the network. \textit{Common Neighbors (CN)} is defined as:
\begin{equation}
    S_{xy}^{CN} = \,\,\mid \Gamma(x) \cap \Gamma(y) \mid.
\end{equation}

We also use two differently normalized versions of the Common Neighbors metric: \textit{Salton Index} normalized by the degree of the nodes
\begin{equation}
    S_{xy}^{Salton} = \frac{S_{xy}^{CN}}{\sqrt{k_x \times k_y}} = \frac{\mid \Gamma(x) \cap \Gamma(y) \mid}{\sqrt{k_x \times k_y}}
\end{equation}
and \textit{Jaccard similarity} normalized by the total size of the node neighborhood:
\begin{equation}
    S_{xy}^{Jaccard} = \frac{S_{xy}^{CN}}{\mid \Gamma(x) \cup \Gamma(y) \mid} = \frac{\mid \Gamma(x) \cap \Gamma(y) \mid}{\mid \Gamma(x) \cup \Gamma(y) \mid}.
\end{equation}

An example of the similarity based on the global topology of the graph is \textit{Katz Index}. This index counts the number of walks (paths that can visit nodes and edges more than once) between considered nodes of increasing length weighted by the coefficient $\beta < 1$. $(A)^k_{xy}$ is the number of walks  of length $k$ between nodes $x$ and $y$ \cite{Newman18}. Katz index is defined as:
\begin{equation}
    S_{xy}^{Katz} = \beta A_{xy} + \beta^2 (A)^2_{xy} + \beta^3 (A)^3_{xy} + \dots.
\end{equation}
This series converges for $\beta$ less than the inverse of the largest eigenvalue of $A$ to the expression:
\begin{equation}
    S_{xy}^{Katz} = ((I - \beta A)^{-1}-I)_{xy},
\end{equation}
where $I$ is an identity matrix.

To analyze the accuracy of each method, we used the traditional hypothesis testing performance metrics. The confusion matrix in Table~\ref{tbl:ConfusionMatrix} introduces four basic metrics: true positive (TP), false negative (FN), false positive (FP) and true negative (TN) that compare edges identified by the manual 'ground truth' solution (true/false) with the edges obtained by a method (positive/negative). Other important metrics include Precision, Recall, Miss Rate, Accuracy and F-measure that are calculated as functions of TP, FN, FP and TN as:
\begin{equation}
    \begin{array}{lcc}
        Precision & = & \frac{TP}{TP + FP}, \\
        Recall & = & \frac{TP}{TP + FN}, \\
        Miss \,\, Rate & = & \frac{FN}{FN + TP}, \\
        Accuracy & = & \frac{TP + TN}{TP + FP + FN + TN}, \\
        F-measure & =& \frac{TP}{TP + \frac{1}{2}(FP + FN)}.
    \end{array}
\end{equation}

\begin{table}[ht]
    \begin{center}
        \begin{tabular}{ c  c  c | c |}
            & & \multicolumn{2}{c}{\textbf{Prediction by a method}} \\
            \cline{3-4}
            & & \multicolumn{1}{|c|}{Predicted} & Not predicted \\
            \cline{2-4}
            \multirow{3}{*}{\rotatebox[origin=c]{90}{\textbf{truth}}} & \multicolumn{1}{|c|}{ } & True Positive (TP). Correct prediction. & False Negative (FN). Type II error. \\
            & \multicolumn{1}{|c|}{Predicted} & Edge repaired by a manual repair & Edge repaired by a manual repair, \\
            & \multicolumn{1}{|c|}{ } & and a method & but not repaired by a method \\
            \cline{2-4}
            \multirow{3}{*}{\rotatebox[origin=c]{90}{\textbf{Ground}}} & \multicolumn{1}{|c|}{ } & False Positive (FP). Type I error & True Negative (TN). Correct rejection. \\
            & \multicolumn{1}{|c|}{Not predicted} & Edge repaired by a method, & Edge not repaired by either \\
            & \multicolumn{1}{|c|}{ } & but not repaired manually & manual repair or a method \\
            \cline{2-4}
        \end{tabular}
    \end{center}
    \caption{The confusion matrix identifies four variables: TP, TN, FP and FN depending on the positive or negative outcome of the predicted result and the ground truth. True and false corresponds to the ground truth and positive and negative corresponds to the prediction.}
    \label{tbl:ConfusionMatrix}
\end{table}

\begin{table}[ht!]
    \begin{center}
        \resizebox{\textwidth}{!}{
            \begin{tabular}{| c | c | c | c | c | c | c | c | c | c | c |} 
                \hline
                Link Prediction Metric & \# of colors & TP & FP & FN & TN & Precision & Recall & Miss Rate & Accuracy & F-Measure \\
                \hline\hline
                \multicolumn{11}{| c |}{\textbf{Forward}} \\
                \hline\hline
                \ref{eq:mainip} ($c_{ij} = 1$) & 9 & 4 & 2 & 4 & 350 & 0.67 & 0.5 & 0.5 & 0.98 & 0.57 \\
                \hline
                \ref{eq:mainip} ($c_{ij} = \frac{1}{max(d_i, d_j)}$) & 10 & 4 & 2 & 4 & 350 & 0.67 & 0.5 & 0.5 & 0.98 & 0.57 \\
                \hline
                \ref{eq:mainip} ($c_{ij} = \frac{1}{d_i + d_j}$) & 9 & 6 & 0 & 2 & 354 & 1 & 0.75 & 0.25 & 0.99 & 0.86 \\
                \hline
                \ref{eq:mainip} ($c_{ij} = \frac{1}{d_id_j}$) & 9 & 6 & 0 & 2 & 354 & 1 & 0.75 & 0.25 & 0.99 & 0.86 \\
                \hline
                Preferential Attachment & 16 & 5 & 1 & 3 & 352 & 0.83 & 0.63 & 0.37 & 0.99 & 0.71 \\
                \hline
                Common Neighbors & 16 & 3 & 3 & 5 & 348 & 0.5 & 0.38 & 0.62 & 0.98 & 0.43 \\
                \hline
                Salton Index & 18 & 0 & 6 & 8 & 342 & 0 & 0 & 1 & 0.96 & 0 \\
                \hline
                Jaccard similarity & 19 & 0 & 6 & 8 & 342 & 0 & 0 & 1 & 0.96 & 0 \\
                \hline
                Katz Index & 16 & 0 & 6 & 8 & 342 & 0 & 0 & 1 & 0.96 & 0 \\
                
                \hline\hline
                \multicolumn{11}{| c |}{\textbf{Backward}} \\
                \hline\hline
                \ref{eq:mainip} ($c_{ij} = 1$) & 11-12 & 3 & 0 & 7 & 707 & 1 & 0.3 & 0.7 & 0.99 & 0.46 \\
                \hline
                \ref{eq:mainip} ($c_{ij} = \frac{1}{max(d_i, d_j)}$) & 12 & 3 & 0 & 7 & 707 & 1 & 0.3 & 0.7 & 0.99 & 0.46 \\
                \hline
                \ref{eq:mainip} ($c_{ij} = \frac{1}{d_i + d_j}$) & 12 & 3 & 0 & 7 & 707 & 1 & 0.3 & 0.7 & 0.99 & 0.46 \\
                \hline
                \ref{eq:mainip} ($c_{ij} = \frac{1}{d_id_j}$) & 12 & 3 & 0 & 7 & 707 & 1 & 0.3 & 0.7 & 0.99 & 0.46 \\
                \hline
                Preferential Attachment & 17 & 1 & 2 & 9 & 703 & 0.33 & 0.1 & 0.9 & 0.98 & 0.15 \\
                \hline
                Common Neighbors & 20 & 0 & 3 & 10 & 701 & 0 & 0 & 1 & 0.98 & 0 \\
                \hline
                Salton Index & 21 & 0 & 3 & 10 & 701 & 0 & 0 & 1 & 0.98 & 0 \\
                \hline
                Jaccard similarity & 20 & 0 & 3 & 10 & 701 & 0 & 0 & 1 & 0.98 & 0 \\
                \hline
                Katz Index & 18 & 0 & 3 & 10 & 701 & 0 & 0 & 1 & 0.98 & 0 \\
                \hline
            \end{tabular}
        }
    \end{center}
    \caption{Performance of the different link prediction methods I.}
    \label{tbl:MethodComparison1}
\end{table}

\begin{table}[ht!]
    \begin{center}
        \resizebox{\textwidth}{!}{
            \begin{tabular}{| c | c | c | c | c | c | c | c | c | c | c |} 
                \hline
                Link Prediction Metric & \# of colors & TP & FP & FN & TN & Precision & Recall & Miss Rate & Accuracy & F-Measure \\
                \hline\hline
                \multicolumn{11}{| c |}{\textbf{Forward}} \\
                \hline\hline
                \ref{eq:mainip} ($c_{ij} = \frac{1}{d_id_j}$) & 9 & 6 & 0 & 2 & 354 & 1 & 0.75 & 0.25 & 0.99 & 0.86 \\
                \hline
                Preferential Attachment & 17 & 8 & 40 & 0 & 316 & 0.17 & 1 & 0 & 0.89 & 0.29 \\
                \hline
                Common Neighbors & 18 & 8 & 151 & 0 & 205 & 0.05 & 1 & 0 & 0.59 & 0.1 \\
                \hline
                Salton Index & 17 & 8 & 176 & 0 & 180 & 0.04 & 1 & 0 & 0.52 & 0.08 \\
                \hline
                Jaccard similarity & 17 & 8 & 191 & 0 & 165 & 0.04 & 1 & 0 & 0.48 & 0.08 \\
                \hline
                Katz Index & 17 & 6 & 101 & 2 & 253 & 0.06 & 0.75 & 0.25 & 0.72 & 0.1 \\
                
                \hline\hline
                \multicolumn{11}{| c |}{\textbf{Backward}} \\
                \hline\hline
                \ref{eq:mainip} ($c_{ij} = \frac{1}{d_id_j}$) & 12 & 3 & 0 & 7 & 707 & 1 & 0.3 & 0.7 & 0.99 & 0.46 \\
                \hline
                Preferential Attachment & 19 & 10 & 77 & 0 & 637 & 0.11 & 1 & 0 & 0.89 & 0.21 \\
                \hline
                Common Neighbors & 20 & 1 & 84 & 9 & 621 & 0.01 & 0.1 & 0.9 & 0.87 & 0.02 \\
                \hline
                Salton Index & 20 & 0 & 5 & 10 & 699 & 0 & 0 & 1 & 0.98 & 0 \\
                \hline
                Jaccard similarity & 19 & 0 & 5 & 10 & 699 & 0 & 0 & 1 & 0.98 & 0 \\
                \hline
                Katz Index & 20 & 0 & 34 & 10 & 670 & 0 & 0 & 1 & 0.94 & 0 \\
                \hline
            \end{tabular}
        }
    \end{center}
    \caption{Performance of the different link prediction methods II.}
    \label{tbl:MethodComparison2}
\end{table}

In this paper, we use two approaches to choose the number of top edges to be repaired. First, as shown in Table~\ref{tbl:MethodComparison1}, we choose it to be equal to the number of edges repaired by the formulation with the coefficient in the objective equal to $c_{ij} = \frac{1}{d_id_j}$ (6 in the forward circuit and 3 in the backward). Second, we use the process  described in Section \ref{section:indices}, apply it to a given reconstruction method and choose the number of edges corresponding to the maximum number of non-trivial colors of the obtained graphs. Table~\ref{tbl:MethodComparison2} shows results obtained by using this approach. Since the adjacency matrix is fairly sparse, the TN value is very high for all the methods. Therefore the best metric for the overall performance assessment is the F-measure that is proportional to the number of edges guessed correctly and inversely proportional to the sum of missed and falsely identified edges. 

The formulation \eqref{eq:mainip} with 
 sum and multiplication objectives, $c_{ij} = \frac{1}{d_i + d_j}$ and $c_{ij} = \frac{1}{d_id_j}$, respectively, outperform all other objectives and link prediction methods from the literature on both graphs. The next best method is the preferential attachment. This likely happened due to the fact that the repaired networks are hub-driven. That is, both networks have two hubs that are connected to most of the other nodes and to each other, therefore edges that are repaired by the preferential attachment are those that connect hubs to nodes with the highest degree. In bigger networks with more than two hubs, preferential attachment is likely to connect pairs of unrelated nodes with high degrees.

Methods based on the different normalizations of the common neighbors metric performed fairly poorly. These methods are not likely to uncover hidden symmetries because of their underlying assumption: nodes that have a lot of common neighbors are likely to be neighbors themselves. The highest ranking edge repaired by the Salton index in the forward circuit is (DB06, VB11). This edge connects two nodes of the same color and breaks the symmetry between them and the rest of the nodes of that color. Hence, this repair made the network less symmetric rather than more symmetric. Consider now two random nodes of the same color. Recall, nodes of the same color have exactly the same number of neighbors (in an unweighted network) of all the other colors. Very often it means having a lot of the same neighbors. Therefore, measures that are based on the number of common neighbors will often rank nodes of the same color as likely to have a connection, which is likely to break some symmetry in the network rather than restoring it. The possible alternative symmetry-restoring assumption may be: nodes that have a lot of common neighbors are likely to have more common neighbors, which will keep the symmetry between the nodes with a lot of common neighbors and increase the symmetry in their neighborhood.

\section{Conclusion}
\label{sec:conclusion}

In summary, we described a method that allows a user to repair a graph to a more symmetric version and compared our results with the manual repair obtained in \cite{Morone19} as well as showing better performance than any of the traditional link prediction methods. We conclude with  observations and ideas for future work.

\begin{enumerate}
    \item The pseudobalanced coloring obtained by solving \eqref{eq:mainip} formulation sometimes is not minimal as the objective minimizes the weighted sum of links added. So if a pseudobalanced $K$-coloring had the same optimal objective as a pseudobalanced $(K+1)$-coloring, the pseudobalanced $K$-coloring could be returned for the $K+1$ case with one nontrivial node made trivial. This occurred five times and these solutions were not chosen. Moreover, our problem had two objectives: (1) find the pseudobalanced $K$-coloring with the most non-trivial nodes, and (2) minimize the weighted sum of links added. Thus, our results can be interpreted as determining the best solution to use along the pareto optimal tradeoff curve between the number of colors, $K$, and the optimal weighted sum of added links. Under this interpretation, the aforementioned balanced $(K+1)$-coloring would not be on the pareto optimal curve as the $K$-coloring would be considered more optimal on both objectives.
    
    \item During preprocessing some of the nodes in the network are assigned fixed colors using constraints of the form Eq.~\eqref{eq:same-color}. As a result, sets of nodes assigned different non-trivial colors in the original partitioning can't be merged together. Therefore, repairs like the pink $S_{12}$ group in Fig.~\ref{Fig:Backward_repairs_2}c are not obtainable by solving \eqref{eq:mainip} and need to be obtained as an additional step. This can be solved by implementing a two-step process as described in steps 2(a) and 2(b) of Section \ref{sec:results}.
    
    \item Removal of edges or reduction of their is not allowed due to the fact that pseudoedges in Definition \ref{def:pseudo} are constrained to non-negative weights on non-edges. This choice is appropriate when studying neural networks of gap junctions in \textit{C. elegans}. The reconstruction of this network is done using images of cross-sectional areas of the worm obtained using electron microscopy \cite{Durbin87}. Each connection is followed from the neuron through all the layers leading to another neuron, therefore it's much more likely to miss a link than to add a non-existent one. When working on a different biological network, the possibility of the removal of edges or edge weight reduction may be required.
    
    \item The two graphs studied in this paper have a high degree of symmetry. We deduced that, for these graphs, the number of non-trivial colors is a good indicator of the best solution and found the best solution using this index. However further investigation on a greater variety of graphs, including those with directed and weighted edges and those of bigger size, is required to make final conclusions. Biological networks present real dynamical systems that need to be able to maintain certain stable synchronization patterns, therefore indices characterizing stability and synchronizability of the cluster synchronization solution predicted by the balanced coloring are likely to give a better indication of ideal repairs, at least in biological networks, which are expected to produce stable cluster synchronization of their units \cite{LeiferPLOS20}.
    
    \item There are two potential ways of how this problem can be formulated: 1 - find the minimal number of edges to add to find a graph with a given number of colors, 2 - find the repair with minimal number of colors given the amount of edges that can be added. Both formulations make sense and both formulations would produce a number of graphs that will need to be analyzed in order to choose the best solution. The number of colors in a graph is no more than $n$, the number of nodes, therefore the 1st formulation will generate no more than $n$ graphs. The number of edges in a graph is no more than $n^2$, therefore the 2nd formulation can generate up to $n^2$ graphs. In the interest of decreasing the search space, we chose the 1st formulation and the 2nd formulation was not explored.
    
    \item Solutions to optimization problems can often be improved by using expert insights about the object of optimization. For example, in Section \ref{sec:comparison_of_methods} we saw that the result obtained with $c_{ij} = 1$ can be improved by using an assumption that edges between nodes with higher degrees are easier to repair by setting $c_{ij} = \frac{1}{d_id_j}$. Additional insight based on the biological considerations can provide further improvements to the result.
    
    \item {\color{red} SBMs have important applications to network reconstruction methods \cite{Guimera09, Peixoto18, Ghasemian20}. In particular, Guimer\`a and Sales-Pardo ~\cite{Guimera09} describe how the reliability of each potentially missing or spurious link can be calculated using graph generative models. As such, a generative model promotes certain topological features in the repaired graph, i.e. if certain features appear more often in generated graphs, the corresponding links are considered more reliable. We speculate that if a generative model for random equitable graphs (as described for example in ref.~\cite{Newman14}) is chosen with the proper set of parameters, the reconstruction will promote the higher degree of symmetry in the reconstructed network. Parameters of such model can be identified from the large scale analysis of the equitable partitions (balanced colorings) in networks performed in ref.~\cite{MoronePNAS20} using the fibrationSymmetries library (available at \url{https://github.com/makselab/fibrationSymmetries}). Additionally, one of the fundamental problems in network reconstruction via SBMs, and also in general, is the choice of the number of edges to be added/removed. We believe that indices derived from synchronizability and stability as discussed in section VIB can provide a novel approach to this problem.}
    
    \item Our method can be applied to directed graphs by modifying the constraints Eq.~\eqref{eq:approx-weight} in \eqref{eq:mainip} suitably to account for the version of the directed balanced coloring problem that is being solved.  In particular, the edge set $E$ must be modified so that only direct edges are used and the number of constraints reduced to reflect whether both directions of the balancing is required. Future work will include a formulation to repair directed networks and its comparison with the quasifibration formalism developed in \cite{Boldi21}.
    
    \item In this paper we applied our method to the unweighted graphs as a simplest case, however this method can be applied to weighted graphs without further modification by relaxing the integrality of the $z_{ij}$ variables. This would change \eqref{eq:mainip} from an integer linear program to a mixed-integer linear program.
    
    \item Our method is reasonably fast for small graphs. In order to be applied to graphs of bigger size, a more computationally efficient methods such as the Bender's decomposition method described in Appendix~\ref{sec:benders} needs to be implemented.
\end{enumerate}

Overall, we presented a formulation of the ORP following the PBC of the graph. Our analysis shows encouraging results for the case of binary unweighted undirected networks as compared to a manually curated graph and other methods for missing link prediction. More work needs to be done to extend the formulation for large scale networks with weighted and directed edges, including negative weights which are important in all biological networks which contain inhibitory interactions.  

\section{Acknowledgements and Data Availability}

We thank F. Operti, C. Smith, I. Stewart, A. Nazerian and G. Verret for many helpful discussions. Funding was provided by NIBIB and NIMH through the NIH BRAIN Initiative Grant R01 EB028157 and NIH grant 1R21EB028489. We would like to thank the UNM Center for Advanced Research Computing, supported in part by the National Science Foundation, for providing the high performance computing resources used in this work. All code and data are available at \url{https://osf.io/prt5g/} and \url{https://github.com/MakseLab/PseudoBalancedColoring}.

\bibliographystyle{apsrev4-2}
\bibliography{thebib}

\clearpage

\section{Appendix}

\subsection{Additional complexity comments}
Here, we show that balanced coloring is hard when the balanced condition is only enforced between nodes of the same color versus the original version where the balanced condition is enforced between nodes of different colors as well. Such a restriction implies the directed-version of the balanced $K$-coloring problem is NP-Hard when only a subset of the constraints are enforced. {\it This further means the balanced coloring problem is easier when additional constraints are added as the directed-version of the balanced $K$-coloring problem is polynomial time solvable.} Formally, we define this variant as follows.
\begin{definition}
Let $K$ be a given positive integer and $G=(V,E)$ a given directed graph. The in-directed self-balanced $K$-coloring problem is to determine whether there exists a partition, $\cC$, of $V$ which satisfies the following. For all $C \in \cC$, and all pairs of distinct nodes $p,q \in C$, 
\begin{equation}
    \label{eq:restricted-balance}
    \sum_{j \in C: (j,p) \in E} A_{jp} =
    \sum_{j \in C: (j,q) \in E} A_{jq}.
\end{equation}
Also, $|\cC|=K$.
\end{definition}
Note how Eq.~\eqref{eq:restricted-balance} is a subset of Eq.~\eqref{eq:balanced-in} restricted to the sum of edge weights within the color set of $p$ and $q$. We show that finding such a coloring is NP-Hard. We show this by reducing {\it traditional vertex $K$-coloring} to restricted directed in-balanced $K$-coloring. Traditional vertex $K$ coloring is the problem of coloring a graph so that no two vertices of the same color share an edge. Formally, we can define it as follows.
\begin{definition}
Let $K$ be a given positive integer and $G=(V,E)$ a given undirected graph with edge weights $w_e$ for $e \in E$. The traditional vertex $K$-coloring problem is to determine whether there exists a partition, $\cC$, of $V$ such that $|\cC|=K$ and for all $C \in \cC$, if $u,v \in C$ then $uv \not\in E$.
\end{definition}
For $K \geq 3$, traditional vertex $K$-coloring is NP-Complete as shown by Karp \cite{Karp72}.
By \red{augmenting the input graph to $K$-vertex coloring with the appropriate subgraph, }we can perform the reduction.
\begin{theorem}
\label{thm:restricted-coloring-NPHard}
Let $G=(V,E)$ and $K \geq 3$ be a given instance of the $K$-vertex coloring problem. The restricted directed in-balanced $K$-coloring problem is NP-Hard.
\end{theorem}
\begin{proof}
Let $G'=(V \cup W, E' \cup F)$ be a directed graph where $V$ is the set of nodes from the $K$-vertex coloring instance, $E'$ is the set of directed edges representing edges in both direction for every edge in $E$, i.e.,
\[
E' = \{ij, ji : ij \in E\},
\]
$W = \{v_1,\ldots,v_k\}$ are $K$ additional nodes, and 
\[
W = \{v_i v_j : i > j\}
\]
are a set of edges so that $v_1$ has $K-1$ edges from nodes $v_2,\ldots,v_k$, $v_2$ has $K-2$ edges, etc. First, note that each of node in $W$ must be assigned a different color or else the balanced condition would be violated. This implies that no node can have any incoming edge from a node of the same color. Thus, the coloring on $V$ must be a $K$-vertex coloring of the original graph.
\end{proof}

\subsection{A Bender's decomposition approach}
\label{sec:benders}

For the size instances we report on, the runtime to solve the integer linear program was acceptable. However, we suspect runtimes will scale poorly on larger instances. Because of this, we describe an adaptation of a Bender's decomposition
approach due to Codato and Fischetti \cite{CF05} that accelerate solution times for integer programs with big-M constraints such as Eq.~\eqref{eq:approx-weight} for future implementation. In this approach, we
define
\begin{equation}
  \label{eq:leader}
  L_\cV = \{\bs = (x,y) \in \{0,1\}^{|V|^2} \times \{0,1\}^{|V| \times K}: \eqref{eq:everycolor}, \eqref{eq:onecolor},
                  \eqref{eq:color-pair}, \eqref{eq:antisymmetry}, \cV \}
\end{equation}
where $\cV$ denotes a set of constraints of form Eq.~\eqref{eq:valid} as defined below.
The leader problem is to find a solution $\bs^* = (x^*,y^*) \in L_\cV$. Given such an $\bs^*$, we calculate 
\begin{equation}
\label{eq:follower-sets}
EK^* = \{(i,j,k): y^*_{jk} = 1, (i,j) \in E^C\} \mbox{  and }
V_2^* = \{(p,q): x^*_{pq} = 1 \}.    
\end{equation} 
Note that $EK^*$ represent the set
of all non-edges that the solution $\bs^*$ permit to exist and that $V_2^*$ is the set of node pairs which must be balanced based on the solution $\bs^*$. Then, given an upper bound $U^*$ on $B^*$, the follower
problem constraints are as follows. We have an objective bound constraint.
\begin{equation}
    \label{eq:follower-objective}
    \sum_{(i,j,k) \in EK^*} z_{ijk} \leq U^* - \delta.
\end{equation}
The constraints in Eq.~\eqref{eq:approx-weight} are modified as $\bs^*$ indicates which nodes must be balanced.
\begin{equation}
    \label{eq:follower-approx-weight}
    \sum_{j:(p,j,k) \in EK^*} z_{pjk}  +\sum_{j\in V: pj \in E} A_{pj}=
    \sum_{j:(q,j,k) \in EK^*} z_{qjk}+\sum_{j\in V: qj \in E} A_{qj}, (p,q) \in V_2^*, k \in K.
\end{equation}
Finally, Eq.~\eqref{eq:repair-edge-equal} are modified to only include the pseudoedges used.
\begin{equation}
  \label{eq:follower-repair-edge-equal}
    \sum_{k:(p,q,k) \in EK^*} z_{pqc} = \sum_{k:(q,p,k) \in EK^*} z_{qpk}, pq \in E'. 
\end{equation}
Then, we define
\begin{equation}
    \label{eq:follower}
    \cF_{\bs^*} = \{z \in \{0,1\}^{|V|^2 \times K} : \eqref{eq:follower-objective}, \eqref{eq:follower-approx-weight},
    \eqref{eq:follower-repair-edge-equal} \}
\end{equation}
and the follower problem is to find a solution in $\cF_{\bs^*}$.
If $\cF_{\bs^*}$ is empty then an {\it Irreducible
  Inconsistent Subsystem (IIS)} is found \cite{VanLoon81}. An IIS is a subset of
the constraints of $\cF_{\bs^*}$ such that
\begin{itemize}
\item the system defined by the subset of constraints are infeasible; and
\item removing any one of the constraints results in a feasible system.
\end{itemize}
Note that the constraints in the IIS are indexed by a union of a subset of
$V_2^* \times K$ and a subset of $E'$. Using the IIS found, let $VK^I_2$ and $E'_I$ denote these subsets, respectively.

Note that each of the constraints in the IIS corresponds to several binary variables in Eq.~\eqref{eq:leader}. First, consider an arbitrary element $(p,q,c) \in VK^I_2$. In this case, the
binary variable $x^*_{pq}=1$ must be true for the equality constraint to be a part of the formulation. In addition, the constraint could be violated some $z_{pjc}$ and $z_{qjc}$ are not in the formulation as $y_{jc} = 0$, i.e., $(p,j,c) \not\in EK^*$ or $(q,j,c) \not\in EK^*$ and $pj \in E'$ or $qj \in E'$.  Let $VK^*_C = \{(j,c): y_{jc} = 0\}$. Then, the indices of the $y_{jc}$ set to zero that could potentially cause the infeasiblity are described by the set $\{(j,c) \in VK^*_C :pj \in E' \mbox{ or } qj \in E'\}$. 

Now consider an arbitrary $pq \in E'_I$. In this case, the cause of the violation must be due to one or more $z_{pjc}$ or $z_{qjc}$ being held to zero. So, again, the set of indices of $y_{jc}$ that could cause the infeasibility is $\{(j,c) \in VK^*_C :pj \in E' \mbox{ or } qj \in E'\}$.

Indices of the binary variables that could be causing the infeasibility are then
\begin{equation}
    \label{eq:infeasibilityIndices}
    \begin{array}{l}
    I_x = \left\{(p,q): \exists c \in K, (p,q,c) \in VK^I_2\right\} \\
    I_y = \left\{(j,c) \in VK^*_C: (pj \in E' \mbox{ or } qj \in E') \mbox{ and }
                            (\exists c \in K, (p,q,c) \in VK^I_2 \mbox{ or }
                            pq \in E'_I)\right\}.
    \end{array}
\end{equation}

The valid inequality that should be added to Eq.~\eqref{eq:leader} is
\begin{equation}
\label{eq:valid}
\sum_{(j,c) \in I_y} (1-y_{jc})+
\sum_{(p,q) \in I_x} x_{pq} \leq 
|I_y| + |I_x| - 1.
\end{equation}
Note that this inequality may not be the strongest possible as the cause of the infeasibility in any particular constraint could be from one or more $z_{pqc}$ variables or, in the case of the constraints in $VK^I_2$, from the setting that $x_{pq}$ is one (which causes the existence of the constraint). Thus, we must have a systematic way of adding the $z_{pqc}$ variables back to Eq.~\eqref{eq:follower}.

 If Eq.~\eqref{eq:follower} is feasible, an improved upperbound can be found by solving the following optimization version of the follower problem.
\begin{equation}
  \label{eq:followeropt}
    U^* = \min\left\{\sum_{ij \not\in E, c \in K} z_{ijc}: \eqref{eq:follower-approx-weight},
    \eqref{eq:follower-repair-edge-equal}\right\}
\end{equation}
Denote the optimal solution to Eq.~\eqref{eq:followeropt}
by $z^*$. The objective value $U^*$ is used to update the upper bound in used in
Eq.~\eqref{eq:follower-objective} and Eq.~\eqref{eq:follower} is then resolved.

The precise description of the method is as follows. We note that
$\delta$ has not been specified and is dependent on the weights of
the edges in the input graph. For an unweighted graph, we can use $\delta=1$.

\begin{algorithm}[H]
\SetAlgoLined
\KwResult{An optimal solution, $(\bs^*,z^*)=(x^*,y^*,z^*)$, to \eqref{eq:mainip}}
  Initialize $\cV$ to the empty set\;
  Use Eq.~\eqref{eq:leader} to form $L_\cV$\;
  Set $U^*$ to an upperbound, e.g., the maximum possible edge weight times $|V|^2$\;
  Initialize $z^*$ to zero.\;
 \While{$L_\cV \not= \emptyset$}{
  Find a feasible solution $\bs^*\in L_\cV$\;
  Use $\bs^*$ to find $EK^*$ and $V_2^*$ via Eq.~\eqref{eq:follower-sets}\;
  Use Eq.~\eqref{eq:follower} to form $\cF_{\bs^*}$\;
  \eIf{$\cF_{\bs^*} \not= \emptyset$}{
    Solve Eq.~\eqref{eq:followeropt} and update $U^*$ and
    $z^*$\;
   }{
   Find an IIS, calculate $I_x$ and $I_y$ via Eq.~\eqref{eq:infeasibilityIndices}\;
   Use $I_x$ and $I_y$ to add a valid inequality to $\cV$ via Eq.~\eqref{eq:valid}\;
  }
 }
 \Return{$(\bs^*,z^*)=(x^*,y^*,z^*)$}
 \caption{Bender's decomposition of \eqref{eq:mainip}}
\end{algorithm}

\end{document}